\documentclass[useAMS,usenatbib,usegraphicx,a4paper]{mn2e}
\usepackage{deluxetable} % this is for mnras
\usepackage[total={17.8cm,24.0cm},centering]{geometry}
\usepackage{times}

\newcommand{\kms}{km~s$^{-1}$}
\newcommand{\kkms}{K~km~s$^{-1}$}
\newcommand{\tastar}{$\rm T_A^*$}
\newcommand\solmass{$\rm M_{\sun}$}
\newcommand\solum{$\rm L_{\sun}$}
\newcommand{\asec}{\arcsec}
\newcommand{\error}{$\pm$}
\newcommand{\lr}{\lambda_R}
\newcommand{\lreps}{\lambda_R/(0.31\sqrt{\epsilon})}
\newcommand{\htoo}{H$_2$}
\newcommand{\mhtoo}{M(H$_2$)}
\newcommand{\mhtoolk}{M(H$_2$)/L$_K$}
\newcommand{\mk}{M$_K$}
\newcommand{\mjam}{M_{\rm JAM}} % needs to be used inside $$

\newcommand{\msunsqpc}{$\rm M_\odot~pc^{-2}$}
\newcommand{\atlas}{{ATLAS$^{\rm 3D}$}}
\newcommand{\sigmae}{$\sigma_e$}

\begin{document}

\title[Molecular gas content of early-type galaxies]{The \atlas\ project -- IV: the
molecular gas content of early-type galaxies\thanks{
Based on observations carried out with the IRAM 30m Telescope. IRAM is
supported by INSU/CNRS (France), MPG (Germany) and IGN (Spain).
}}
\author[L.\ M.\ Young et al.]{Lisa M.\ Young$^{1,2}$,\thanks{E-mail: lyoung@physics.nmt.edu} 
 Martin Bureau$^3$,
 Timothy A.\ Davis$^3$,  
 Francoise Combes$^4$,
 \newauthor
 Richard M. McDermid$^5$,
 Katherine Alatalo$^6$, 
 Leo Blitz$^6$,
 Maxime Bois$^{7,8}$,
 \newauthor
 Fr\'ed\'eric Bournaud$^{9}$,
 Michele Cappellari$^3$,
 Roger L. Davies$^3$,
 P. T. de Zeeuw$^{7,10}$,
 \newauthor
 Eric Emsellem$^{7,8}$,
 Sadegh Khochfar$^{11}$,
 Davor Krajnovi\'c$^7$,
 Harald Kuntschner$^7$,
  \newauthor
 Pierre-Yves Lablanche$^{8}$,
 Raffaella Morganti$^{12,13}$,
  Thorsten Naab$^{14}$,
  Tom Oosterloo$^{12,13}$,
  \newauthor
  Marc Sarzi$^{15}$,
  Nicholas Scott$^3$,
  Paolo Serra$^{12}$,
  and Anne-Marie Weijmans$^{16,17}$\\
  $^1$Physics Department, New Mexico Institute of Mining and Technology,
 Socorro, NM~87801, USA\\
  $^2$Adjunct Astronomer, National Radio Astronomy Observatory, Socorro, NM 87801,
USA\\
  $^3$Sub-department of Astrophysics, Department of Physics, University of Oxford, Denys Wilkinson Building,
Keble Road, Oxford, OX1 3RH, UK\\
  $^4$Observatoire de Paris, LERMA, 61 Av. de l'Observatoire, 75014 Paris, France\\ 
  $^5$Gemini Observatory, Northern Operations Center, 670 N. A$^\prime$ohoku Place, Hilo,
HI 96720, USA\\
  $^6$Department of Astronomy, Campbell Hall, University of California,
Berkeley, CA 94720, USA\\
  $^7$European Southern Observatory, Karl-Schwarzschild-Str. 2, 85748 Garching,
Germany\\
  $^8$Universit\'e Lyon 1, Observatoire de Lyon, Centre de Recherche Astrophysique de
Lyon and Ecole Normale Sup\'erieure de Lyon, \\
~~ 9 avenue Charles Andr\'e, F-69230 Saint-Genis Laval, France\\
  $^9$Laboratoire AIM Paris-Saclay, CEA/IRFU/SAp – CNRS – Universit\'e Paris Diderot,
91191 Gif-sur-Yvette Cedex, France\\
  $^{10}$Sterrewacht Leiden, Leiden University, Postbus 9513, 2300 RA Leiden, the
Netherlands\\
  $^{11}$Max-Planck-Institut f\"ur extraterrestrische Physik, PO Box 1312, D-85478
Garching, Germany\\
  $^{12}$Netherlands Institute for Research in Astronomy (ASTRON), Postbus 2, 7990
AA Dwingeloo, the Netherlands\\
  $^{13}$Kapteyn Astronomical Institute, University of Groningen, Postbus 800, 9700 AV
Groningen, The Netherlands\\
  $^{14}$Max-Planck-Institut f\"ur Astrophysik, Karl-Schwarzschild-Str. 1, 85741 Garching,
Germany\\
  $^{15}$Centre for Astrophysics Research, University of Hertfordshire, Hatfield,
Herts AL1 09AB, UK\\
  $^{16}$Dunlap Institute for Astronomy \& Astrophysics, University of Toronto, 50 St. George Street,
Toronto, ON M5S 3H4, Canada\\
  $^{17}$Dunlap Fellow\\
  }
\maketitle

\begin{abstract}
We have carried out a survey for CO J=1-0 and J=2-1 emission in the
260 early-type galaxies of the volume-limited \atlas\ sample,
with the goal of connecting their star formation and assembly histories to their
cold gas content.
This is the largest volume-limited CO survey of its kind and is the first to
include many Virgo Cluster members.  
Sample members are dynamically hot galaxies with a median stellar mass $\sim
3\times 10^{10}$ \solmass; they are selected by morphology rather than colour, and
the bulk of them lie on the red sequence.
The
overall CO detection rate is 56/259 = 0.22 \error 0.03, with no dependence on $K$
luminosity and only a modest dependence on dynamical mass.
There are a dozen CO detections among the Virgo Cluster members; statistical analysis of
their \htoo\ mass distributions and their dynamical status within the cluster shows that
the cluster's influence on their molecular masses is subtle at best, even though 
(unlike spirals) they
seem to be virialized within the cluster.  We suggest that the cluster members have
retained their molecular gas through several Gyr residences in the cluster. 
There are also a few extremely CO-rich early-type galaxies with \htoo\ masses
$\ga 10^9$ \solmass, and these are in low density environments.
We do find a significant trend between molecular content and the stellar specific
angular momentum.  The galaxies of low angular momentum also have low
CO detection rates, suggesting that their formation processes were more effective
at destroying molecular gas or preventing its re-accretion. 
We
speculate on the implications of these data for the formation of various sub-classes
of early-type galaxies.
\end{abstract}
 
 
\section{INTRODUCTION}
\label{sec:intro}
 
Understanding galaxy formation and evolution is at the
heart of much of current astrophysics.  The task is, however, greatly hindered by 
the lack of quantitative, physically-driven (as opposed to empirical) models of the
regulation of molecular gas and star formation on galactic scales.
All galaxies or protogalaxies must have begun as gas-rich entities, and during
their first epochs of star formation would have belonged to the 
``blue cloud" in a galaxy colour-magnitude diagram.
In contrast, today there is a clear separation between the blue cloud (composed mainly
of star-forming disc and dwarf galaxies) and the red
sequence (old and dynamically hot stellar
systems).  Thus, an outstanding question in galaxy evolution is how the present-day
early-type galaxies moved quickly from the blue cloud to the red sequence.

This rapid movement to the red sequence requires, at minimum, an abrupt cessation of star formation at
high redshift so that the
global colours change from blue to red \citep{thomas05,thomas10}.  
Since today's early-type galaxies are generally rather poor in atomic and
molecular gas \citep{lkrp91}, it is often assumed that the cessation of 
star formation was
achieved by removing, destroying, or consuming the cold gas. 
The most massive early-type galaxies are often rich in hot gas
\citep{osullivan}, so it is possible that molecular gas is destroyed by heating.
An intriguing alternative suggestion is that the cold gas might be retained but
rendered unsuitable for star formation activity \citep{martig}.
Thus, the story of the
development of the red sequence is, to a large extent, a story of what happens to the cold gas
in galaxies.

Several mechanisms have been proposed to deplete the cold gas from 
galaxies and move them onto the red sequence.  
For example, a major merger between two gas-rich galaxies could (depending on
geometry) result in a large-scale loss of angular momentum, dropping gas to the
center of the merger remnant \citep{barnes02}.  The gas could then be funneled into a black hole
and/or consumed in a LIRG-type burst of star formation activity. 
Alternatively, the energy input into the interstellar medium from an active
nucleus or a starburst could hypothetically destroy and/or unbind the cold gas, as
well as prevent hot gas from cooling. 
Environmental effects such as ram pressure stripping, gravitational interactions and galaxy harassment could
also be important in at least some early-type galaxies.  
In fact, it is likely
that all of these processes are important at some level, but we do not yet have
strong constraints on their relative impacts.  The behavior of the cold gas in all
of these transformational processes needs to be better understood.

It has also been known for some years that today's early-type galaxies, while
relatively poor in cold gas compared to spirals, are not completely devoid of cold
gas.   The gas is detected in dusty silhouette discs \citep[e.g.][]{goudfrooij94}, HI
emission \citep[e.g.][]{wardle86,hucht95,moetal06,oo10}, and CO emission
\citep{lkrp91,wch95,CYB,WSY10}.  The origin of this gas is not well understood; it could have
come from internal stellar mass loss \citep[e.g.][]{MB03,Ci10}, or it could have been
acquired from an external source such as another galaxy or cold mode accretion. 
In some early-type galaxies the case for an external origin is very clear because
the specific angular momentum of the gas is dramatically different from that of
the stars.  Cen~A \citep{Quillen} and NGC\,3032 \citep{YBC} exhibit this property,
but in many other cases the gas kinematics are consistent with an internal origin.
Again, probably both external and internal processes contribute at some level.  A quantitative
understanding of their contributions would make it possible for the gas to serve
as an indicator of the galaxies' evolutionary histories and the processes that
moved these galaxies onto the red sequence.

These considerations motivate a study of correlations between the cold gas
content of early-type galaxies and their other properties.  
For example, the
kinematic structures in early-type galaxies (dynamically cold stellar discs,
counterrotating cores, or even the absence of a measurable net rotation) 
preserve clues to their assembly histories.  If the cold gas in these galaxies is
leftover from the time of their assembly, we might expect the gas content to be
correlated with kinematic structure.  In addition, it would be useful to test the
influence of AGN activity and/or local environment on the gas content.
Looking to the future, observations of the cold gas in early-type galaxies can also be
used for testing theoretical models of the ISM structure and of star formation
processes.

We address these goals through a CO survey of the \atlas\ sample,  
a large and complete volume-limited sample of
early-type galaxies spanning a variety of environments and AGN activity levels. 
This is the largest survey to date of CO in early-type galaxies, 
being factors of 5 to 10 larger than previous studies with similar sensitivity.
It is also the first
volume-limited sample to have spatially resolved stellar kinematics, stellar populations,
and ionized gas data for all target members.  And while the smaller CO survey of
the SAURON early-type galaxies \citep{CYB} included both Virgo Cluster members and
field galaxies, it was not a volume-limited sample.  The present paper is the first
to allow reliable statistical analyses of gas contents inside and outside the Virgo
Cluster.

Section \ref{sec:sample} summarizes the selection criteria for the \atlas\
sample.  Section \ref{sec:data} describes the 204 new CO observations obtained for
this survey and the literature data collected for the remainder.
Section \ref{sec:results} describes global results of the survey, and section
\ref{sec:correl} analyzes the molecular gas content compared to a variety of galaxy
properties including stellar kinematics and
environment.  Correlations with stellar populations, star formation activity, and AGN
activity will be discussed in a future paper. 
The discussion in section \ref{sec:discussion} covers implications for the origin of
the molecular gas in early-type galaxies, the removal of gas in clusters and the possible
transformations of spirals into early-type galaxies.

\section{SAMPLE}\label{sec:sample}

The \atlas\ sample is a complete volume-limited sample of early-type galaxies brighter than
$M_K = -21.5,$ covering distances out to 42 Mpc, with some
restrictions on Declination and Galactic latitude \citep[see][hereafter Paper I]{paper1}.
The early-type sample is actually drawn from a parent sample which has no
colour or morphological selection, and optical images of the entire parent sample
have been inspected by eye for large-scale spiral structure.  
The presence or absence of spiral structure is, of course, a physically motivated
selection criterion because it tests whether the stellar population is dynamically
cold (and self-gravitating) or not.
The 260 galaxies lacking spiral structure form the basis of the \atlas\ project.
Integral-field optical spectroscopy was obtained with the SAURON
instrument on the William Herschel Telescope; 
it covers
a field of at least 33\asec $\times$ 41\asec, typically extending to one effective
(half-light) radius $R_e$. 

Paper I gives additional information on the details of the morphological selection of
the sample, assessment of Virgo Cluster membership, the
distances of the galaxies and measured optical velocities.
Paper I shows that the morphologically-selected \atlas\ galaxies
clearly trace out the red sequence, with a small population of bluer
galaxies in the ``transition region" between the blue cloud and the red sequence.
\citet{Davor} (Paper II) present a morphological classification of the internal
stellar kinematic structure of the galaxies, and
\citet{eric} (Paper III) analyse the stellar specific angular momentum 
proxy $\lr$ for the galaxies. 
\citet{P7} (Paper VII) present measurements of the local galaxy density, which we use
for study of environmental effects on cold gas content.

\section{DATA}
\label{sec:data}

\subsection{IRAM 30~m observations and data reduction}
\label{sec:obs}

The IRAM 30~m telescope at
Pico Veleta, Spain, was used for simultaneous observations of $^{12}$CO (1-0) and
(2-1) during July 2007, March 2008, and November 2008. The beam FWHM is
respectively $21.6\arcsec$ and $10.8\arcsec$ at the two frequencies. The
SIS receivers were used for observations in the wobbler switching mode, with reference
positions offset by $\pm 100\arcsec$ in azimuth.  The $1$~MHz
filterbank back-end gave an effective total bandwidth of $512$~MHz (1330 \kms)
and a raw spectral resolution of 2.6 \kms\ for CO(1-0).
The $4$~MHz filterbank gave an effective total bandwidth of
$1024$~MHz (also 1330 \kms) and a raw spectral resolution of 5.2 \kms\ for CO(2-1).

If the CO emission is quite extended,
the relatively small wobbler throw could potentially result in some emission being
located in the ``off" positions.   The signature of this problem would be a
negative feature, most likely at a velocity different to the optical systemic
velocity.  There is no strong evidence of this problem in any spectra except for
those of NGC\,4649, where known CO emission from NGC\,4647 \citep{young06} appears as a negative feature
offset by 270 \kms.  The effect may possibly be seen in the spectra of 
NGC\,2685, which however is already known to have extended CO emission (see below).

The system temperatures
ranged between $190$~K and $420$~K at $2.6$~mm, and between $240$~K and
$600$~K at $1.3$~mm.  A small number of targets had significantly poorer
quality data, with system temperatures up to 1400 K for CO(2-1), and for NGC\,2698
the system temperatures were so large at 2-1 as to make those data useless. 
The pointing was checked every $2$ to 3 hours on a
nearby planet or bright quasar, and focus was checked at the beginning of each
night as well as after sunrise or more often if a suitable planet was available. 
The time on source typically ranged from 12 minutes to 36 minutes and occasionally
longer, being weather dependent, interactively adjusted so that the final co-added
$^{12}$CO (1-0) spectrum for each galaxy had a rms noise level near 3.0 mK
(\tastar) $\approx$ 18 mJy after binning
to 31 \kms\ channels.

The individual 6 minute scans were all inspected and those with unusually poor quality
baselines or other problems were discarded, to reduce the possibility that a faint emission line
feature could be created by baseline instabilities.   The good scans were
averaged together, weighted by the square of the system temperature.
In six cases the CO(2-1) spectrum required a linear (first order) baseline, while in all
other cases a simple constant (zero order) baseline was used.  If no emission was apparent in
the spectrum, all of the channels were used to determine the baseline level.
If emission was obvious the baseline was determined using interactively
selected line-free regions of the spectrum.

Integrated intensities for each galaxy were computed by summing the spectrum 
over velocity.  In the cases with weak or absent line emission, the velocity range is 
300 \kms\ centered on the systemic velocity of stellar absorption lines as measured
from our SAURON data \citep{paper1}.   In most cases our new velocities agree with
values listed in the HyperLEDA database.  For NGC\,4486A our new velocity 
disagrees by 700 \kms\ due to the presence of a bright foreground star corrupting
previous long-slit spectroscopy, and our original use of the LEDA velocity means that 
we have effectively no CO data for that galaxy.
For obvious detections the integrated velocity range covers the real emission.
Table
\ref{tab:intensities} lists the rms noise levels in each spectrum, the velocity ranges
summed, and the integrated intensities $I_{1-0}$ and $I_{2-1}$. 
As in \citet{SWY1}, the
statistical uncertainty $\sigma_I$ in a sum over $N_l$ channels, each of width $\Delta v$ and
rms noise level $\sigma$, is 
\begin{equation}
\sigma_I^2 = (\Delta v)^2 \sigma^2 N_l (1 + N_l/N_b).
\label{noiseformula}
\end{equation}
A contribution from the uncertainty in estimating the baseline level is
included ($N_l/N_b$ is the ratio of the number of channels in the line to the
number of channels used for measuring the baseline level).  
When the integrated intensity is
greater than three times its own uncertainty the galaxy is counted as a detection.
Detected lines were also fit with Gaussian or double-horned template profiles.  The
fitted line areas are consistent with the summed areas;
specifically, for 30 galaxies with 1-0 integrated intensities $> 3$~\kkms , the mean
ratio (sum area)/(fit area) is 0.96 with a dispersion of 0.04.  
The properties of the line fits are discussed in greater detail in section
~\ref{sec:distrib} below.

Figure \ref{fig:detectionstats} shows the integrated
intensities of all sample galaxies, in units of their own statistical uncertainty
(equation \ref{noiseformula}).  The long positive
tail represents real detections; the peak symmetric about 0.0 is caused by
thermal noise and low level baseline wiggles in the spectra.  
In general, one would expect the statistical uncertainty of equation
\ref{noiseformula} to
underestimate the true uncertainty in a sum because it does not account for
correlated channels (baseline ripples), but Figure \ref{fig:detectionstats}
suggests this effect is small since the dispersion in the integrated areas is not
substantially larger than the theoretical estimate.
In our sample there are
a total of 56 galaxies with line intensities above the 3$\sigma$ level; four of them
have $3.0 \leq I_{1-0}/\sigma_I \leq 4.0$ and might be treated with
caution.  Of these four, NGC\,4684 is also detected
independently in the CO(2-1) line at a higher level of significance.
In 203 empty spectra we would 
expect to find only 0.26 false positives with $I/\sigma_I > 3.0$ and also 0.26 
false negative
``detections" (noise that looks like an absorption line) with $I/\sigma_I < -3.0$.  As usual, the real outlier detection rates
are probably larger than that, but the fact that we find so few $3\sigma$ 
negative ``detections" (Figure \ref{fig:detectionstats}) suggests that there are
also few false positives.

\begin{figure}  % 1: histogram of I10
\includegraphics[width=8cm,trim=1cm 0.7cm 1cm 0cm]{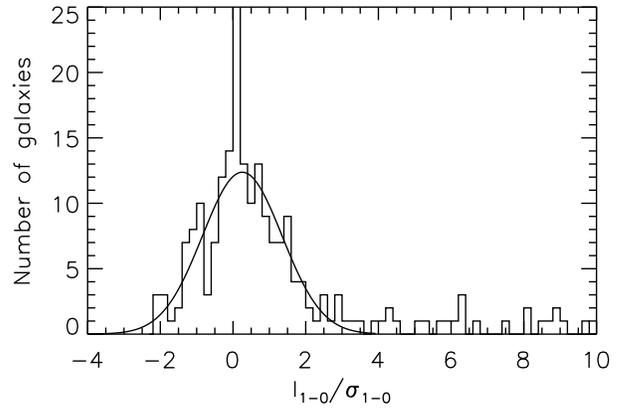}
\caption[]{Integrated intensities.  For each target, the integral of the CO(1-0)
line over the relevant spectral window (300 \kms\ or the detected emission) is
plotted in units of its own statistical uncertainty.  Detected galaxies have a
long tail up to $I_{1-0}/\sigma_{1-0} = 42,$ outside the range of this figure. A
Gaussian with a dispersion of $I_{1-0}/\sigma_{1-0} = 1.09$ is overlaid.}
\label{fig:detectionstats}
\end{figure}

Conversion to main beam brightness temperatures is achieved by dividing the
antenna temperatures \tastar\ by the ratio of the beam and forward efficiencies, 
$B_{\rm eff}/F_{\rm eff}=0.78$ at $2.6$~mm and $0.63$ at $1.3$~mm.
Temperatures quoted in the tables are main beam brightness temperatures
($T_{mb}$).
The conversion from main beam temperature to flux density in Jy is 
4.73 $\rm Jy~K^{-1}$ for the 30m at both frequencies.
The total mass of molecular hydrogen is estimated using a ratio
N(H$_2$)/I$_{1-0}$~$=3\times10^{20}$~cm$^{-2}$
(K~km~s$^{-1}$)$^{-1}$ \citep{dickman86,strong88,strong04}.  This ratio converts an 
integrated intensity to an \htoo\ column density averaged over the beam, and
multiplication by the beam area gives the total \htoo\ mass as
\[
\rm  \frac{M(H_2)}{M_\odot} = 6.0\times 10^4\; \biggl(\frac{D}{Mpc}\biggr)^2 
  \biggl( \frac{\int{T_{mb}\; dv}}{K\;km\;s^{-1}} \biggr). 
\]
A beam size of 21.8\asec\ is assumed and the effective
beam area is $1.13\, \theta_{\rm FWHM}^2$.
The molecular mass thus only refers to the mass within the 
central 22\asec\ of the galaxy, which corresponds to a diameter of 2.2 kpc at a
typical distance of 20 Mpc.
The masses will be underestimated if the molecular gas is more
extended than the beam of the 30m telescope.  Based on previous experience with
interferometric imaging of CO in early-type galaxies, this effect can occasionally be
as large as a factor of two or more
\citep[e.g.][]{y02,y05,YBC} but it cannot be predicted {\it a priori}.
The beam of the 30m telescope for CO(1-0) does cover roughly one third of the {\tt
SAURON} field-of-view ($33\arcsec\times41\arcsec$ in low-resolution
mode), so in most cases the mean stellar parameters are determined over a region of
similar size to the CO(1-0) beam.
Interferometric observations of the CO detections are being made with the CARMA
and Plateau de Bure arrays, and these will find molecular gas which is more
extended than the 30m beam.

\subsection{Duplicate observations}

NGC\,3412, NGC\,3941, NGC\,4026, and NGC\,4346 were observed for this project and were
also observed with the 30m telescope by \citet{WS03}.  The two sets of 
observations have similar noise levels and the results in all four cases
are consistent between the two papers. NGC\,3941, NGC\,4026, and NGC\,4346 are non-detections
in both papers.  In the case of NGC\,3412, \citet{WS03} claim a detection in 
CO(1-0) at about the $3.8\sigma$ significance level. 
We find no detection but have somewhat worse noise so
that the flux upper limit we claim is $\approx$ 50\% higher than the flux
they quote.  
Similarly, NGC\,7332 and NGC\,7457 were also observed by \citet{WS03} although using
the
NRAO 12m telescope, and the non-detections reported here are consistent with those
data (our sensitivity is worse for NGC\,7457).

NGC\,3193, NGC\,4494, and NGC\,4697 have CO spectra from the 30m telescope both in 
this paper and in \citet{SWY1}, with similar noise levels.
NGC\,821, NGC\,3226, NGC\,3377, NGC\,3379, NGC\,3640, NGC\,4697, and NGC\,5845 have CO spectra
published both in this paper and in \citet{WSY10}.
Seven of these 10 (all except NGC\,3640, NGC\,4494 and NGC\,3226)
are non-detections in both datasets.  For NGC\,4494, \citet{SWY1} claim a detection
at about the 3$\sigma$ level in CO(1-0); the feature is weak and broad, and
is not duplicated in their 2-1 spectrum.  We do not find any suggestive
features in our spectrum and so we claim a non-detection at a flux level about
30\% smaller than the detection claimed by \citet{SWY1}.  For NGC\,3640,
\citet{WSY10} find a 3$\sigma$ detection in CO(2-1) but not in CO(1-0); we
find no emission in somewhat more sensitive data.
For NGC\,3226, \citet{WSY10} again claim a 3$\sigma$ detection in CO(1-0) at a velocity
of 1100 \kms; however, our new optical spectroscopy indicates the
stellar systemic velocity to be 1321 \kms\ \citet{paper1} and our formal sum over 300 \kms\ centered
on the optical velocity finds no detection.
Given the nature of
$3\sigma$ detections, there is no glaring inconsistency between these
observations.

As mentioned in \citet{CYB}, there are a small number of detections which are
observed by two teams.  NGC\,4150 is detected both by \citet{CYB} and by
\citet{WS03} using the 30m; the CO(1-0) fluxes are entirely consistent, but the
CO(2-1) fluxes differ
by a factor of two.  The difference could easily arise from the telescope
pointing, as \citet{YBC} have shown that the CO emission in NGC\,4150 is compact
and a small pointing offset could mean that much of it is observed at a lower gain.
NGC\,3489 shows a factor of two difference in both CO lines between those 
papers, which could again be a pointing problem or inaccurate spectral baselines.
Finally, it is worth remembering that the absolute calibration of these millimetre
observations is not normally better than 10\%--20\%.
These comparisons suggest that in the fainter detections the 
flux levels (hence \htoo\ masses) might really be uncertain by
up to 50\%.   This is especially true for the CO(2-1) line where the telescope pointing becomes 
both more critical and more difficult because of the smaller beam.  

Eleven of the \atlas\ members have also been observed with the BIMA, CARMA, and
Plateau de Bure interferometers.  Comparisons of the single dish and
interferometer total flux measurements for NGC\,3032, NGC\,4150, NGC\,4459 and
NGC\,4526 are presented by \citet{YBC}; for NGC\,2768 by \citet{crocker2768}; and
for NGC\,4550 by \citet{crocker4550}.   Recent observations of NGC\,524, NGC\,3489,
NGC\,4477, and NGC\,7457 at the Plateau de Bure are reported by
\citet{crocker-all}.  In general there is reasonable agreement,
given the cautionary note mentioned above.  Our 30m nondetection of
NGC\,7457 is confirmed at the Plateau de Bure.
NGC\,4550, which is a non-detection here, does
have a small amount of molecular gas that is quite compact and more easily
detected by an interferometer with a smaller synthesized beam.  
The interferometric flux of NGC\,3489 is larger than what was found with the 30m
because the molecular gas is extended over an area somewhat larger than the 30m
beam.
Other detections are consistent in total flux and linewidth between the single dish
and the interferometer.

\subsection{Literature Data}

NGC\,3607, NGC\,4111, NGC\,4143, NGC\,4203, NGC\,4251, and NGC\,5866 were observed with
the 30m telescope by
\citet{WS03}.  We use molecular masses derived from their reported CO(1-0) line
intensities, scaled to the conversion factor and the distance that we assume.
The method used by \citet{WS03} for estimating uncertainties and upper limits is
similar to the method used here, with the slight difference that the velocity
range used for summing varies between 200 and 400 \kms\ instead of our
constant 300 \kms.
Similarly, NGC\,3073, NGC\,3605, NGC\,4283, and NGC\,4636 were observed by
\citet{SWY1} and we take results from that paper.  The CO emission in NGC\,5866 and
NGC\,3607 seems to be extended, as the line is detected in pointings offset from
the galaxy nucleus.  However, for consistency with the rest of our observations,
only the data from the central pointings are used here.
Data for NGC\,3522 are taken from \citet{WSY10}.

The most recent single dish CO spectrum of NGC\,4476 is published by \citet{wch95}, but
the signal-to-noise ratio in that spectrum is quite low.   We consider the
CO(1-0) flux measured by \citet{y02} in a map made with the BIMA array to be
more reliable and have used that value.

NGC\,4278 was observed at the IRAM 30m telescope
by both \citet{CYB} and \citet{WSY10}.  The first paper reports a tentative detection
of 2.1\error 0.41 \kkms\ in the 1-0 line, whereas the second reports a $3\sigma$ upper
limit of 1.9 \kkms.  An observation with the IRAM Plateau de Bure interferometer 
reports a more sensitive upper limit \citep{crocker-all}.  
The nondetection of \citet{WSY10} is used here.

\section{RESULTS}
\label{sec:results}
 
\subsection{CO detection rate}
\label{sec:incidence}
 
The derived H$_2$ masses and limits for all \atlas\ sample members are listed in 
the last column of Table~\ref{tab:intensities}.
For non-detected galaxies, the quoted
mass limits correspond to three times the statistical uncertainty in
an integral over the assumed line width.
There are 56 detections out of 259 sample members with CO data, giving
an overall detection rate of 0.22 \error\ 0.03.  The quoted error is a $1\sigma$
formal uncertainty assuming a binomial distribution.
Thus, the current detection rate is consistent with the value 0.28 \error 0.08
reported by \citet{CYB} for the much smaller and somewhat closer SAURON sample.
Of the 56 total detections, 46 are detected at greater than $3\sigma$ in
both the CO(1-0) and CO(2-1) lines.  Six additional galaxies (NGC\,0509, NGC\,2685,
NGC\,3245,
NGC\,4283, NGC\,4476, and NGC\,6798) are detected in only the
CO(1-0) line, though NGC\,4476 does not have high quality data in the 2-1 line.
Three galaxies (NGC\,3599, NGC\,4036, and NGC\,4643) are formally detected
only in the CO(2-1) line.

The detection rate reported here is a strict
lower limit to the true incidence of molecular gas in early-type galaxies.
The case of NGC\,4550 was already mentioned above.
In addition, NGC\,2685 (detected here in the 1-0 line but not 2-1) is known to have a significant amount of
molecular gas that is located outside of our central pointing position
\citep{ss02}.
For the sake of consistency and homogeneity, we use our own
data for these two galaxies in the current analysis and simply bear in mind that we have
not yet found all of the molecular gas in early-type galaxies.
Due to the finite sensitivity of the observations there is, of course, also 
a detection bias against broad and weak lines.  

Our current detection rate is strikingly inconsistent with the 0.78 \error\ 0.17
detection rate quoted by \citet{WS03} for a volume-limited sample of S0 galaxies.
Even if we exclude from our \atlas\ sample the 25\% with types $T <
-3.5$ (ellipticals, as classified in LEDA), our detection rate is still only 0.28 \error\ 0.03.
The sample of \citet{WS03} has a factor of 10 fewer galaxies than are presented here,
but the reasons for the difference in detection rate are still not obvious.
\citet{WS03} exclude known Virgo Cluster members, though as we will show later, our
detection rate is not much lower in the Virgo Cluster.
We will also show that our detection rate is not a strong function of stellar
luminosity
over the range covered here ($-26 < M_K < -21.5$).  
Thus, other factors such as selection criteria and sensitivities may contribute to the difference.
The luminosity selection of \citet{WS03} is done on a $B$ luminosity, which means
that at the faint end of the sample the members will be preferentially bluer than
other galaxies of similar stellar mass.  They could thus have larger molecular
content per unit stellar mass.
And as we have mentioned above, it is clear that greater sensitivity would produce a
greater detection rate.
\citet{WS03} have integrated down to a mass limit which
is a fixed fraction of the optical luminosity of the galaxy, and in some cases
this limit is several times deeper than we have been able to achieve.  On the other
hand, our limits are deeper for other objects especially at high luminosities.
Since the integrated CO properties are not trivially related to the optical
properties \citep{CYB}, it is not clear which of the two strategies (fixed thermal
noise level or fixed CO/$L_K$ ratio) would be expected to produce a greater detection
rate.

\subsection{Distribution and temperature of molecular gas}
\label{sec:distrib}

Figure~\ref{fig:spectra1} shows the 30m spectra obtained in the
CO(1-0) and CO(2-1) lines towards the detected galaxies, along with their optical
velocities. 
The detected lines were fit with a Gaussian to provide estimates of the peak intensity, central
velocity, and FWHM (Table~\ref{tab:detections}).   In a few cases the detected line is obviously
better described by a double-horned spectrum than by a Gaussian, and these were
fit with a template which uses a parabolic shape between two sharp line edges.
Comparisons of Gaussian and double-horned fits revealed that the two shapes gave
almost identical estimates for the CO systemic velocity and line width, so those
parameters are robust to the details of the line shape in these cases of moderate
signal-to-noise ratio.
In many other cases the line is too asymmetric to be well fit by either of these
parametrizations, and in these cases the peak intensity, central velocity, and
FWHM are measured directly from the spectra.

\begin{figure*}  % spectra
\begin{center}
\includegraphics[width=16cm,trim=1cm 0.5cm 1cm 1cm]{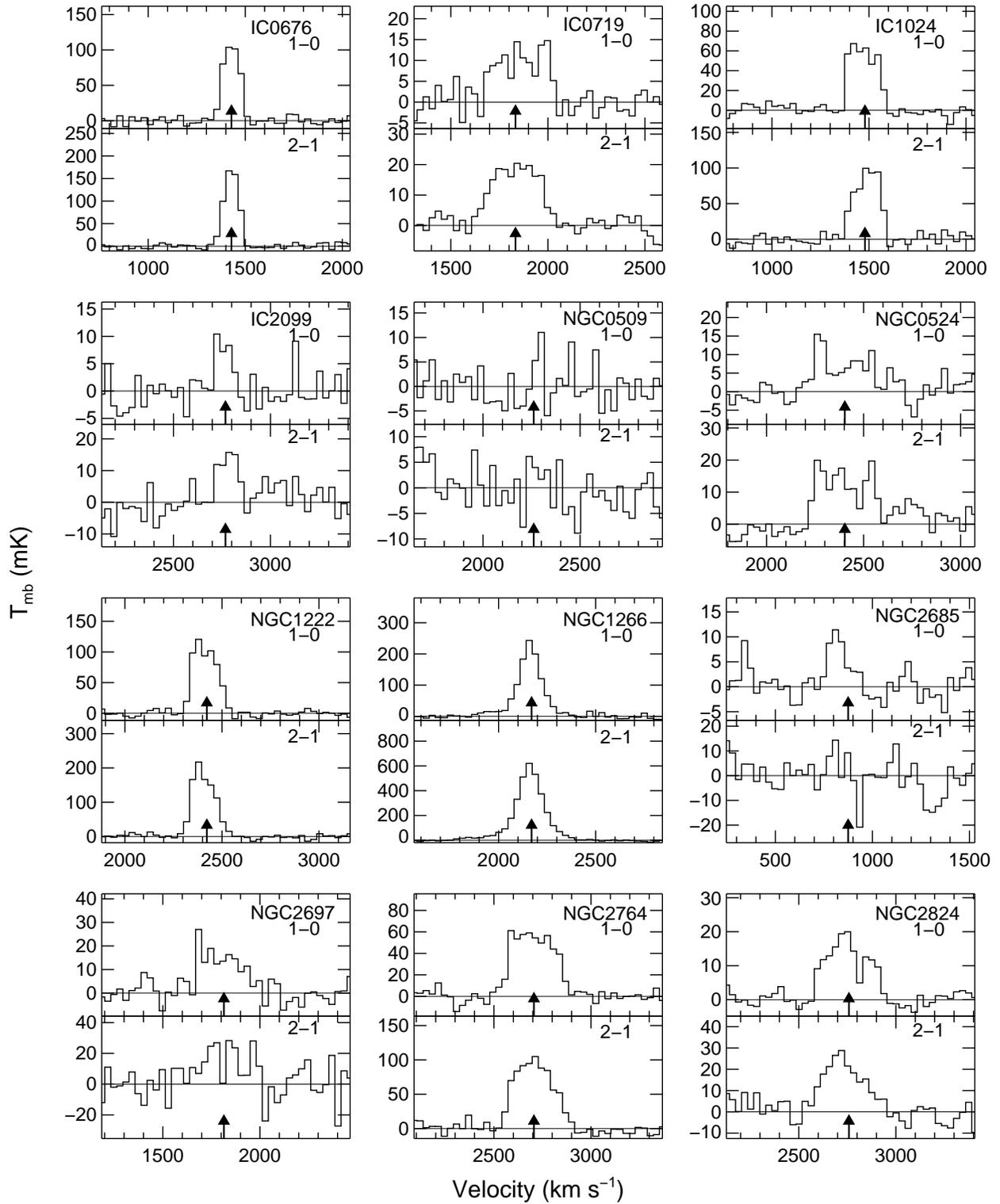}
\caption[]{CO(1-0) and CO(2-1) IRAM 30~m spectra of the galaxies
detected. The spectra have been binned to $30.6$~km~s$^{-1}$ and the
scale is $T_{\rm mb}$ in mK. The CO(1-0) spectra are in the top half of each
panel and the CO(2-1) spectra are in the bottom half.  The arrow indicates the
systemic velocity of stellar absorption lines, taken from our new SAURON data 
\citep{paper1} or from LEDA (for the 7 galaxies observed in CO but dropped from the
\atlas\ sample).}
\label{fig:spectra1}
\end{center}
\end{figure*}

\begin{figure*}
\begin{center}
\includegraphics[width=16cm,trim=1cm 0.7cm 1cm 1cm]{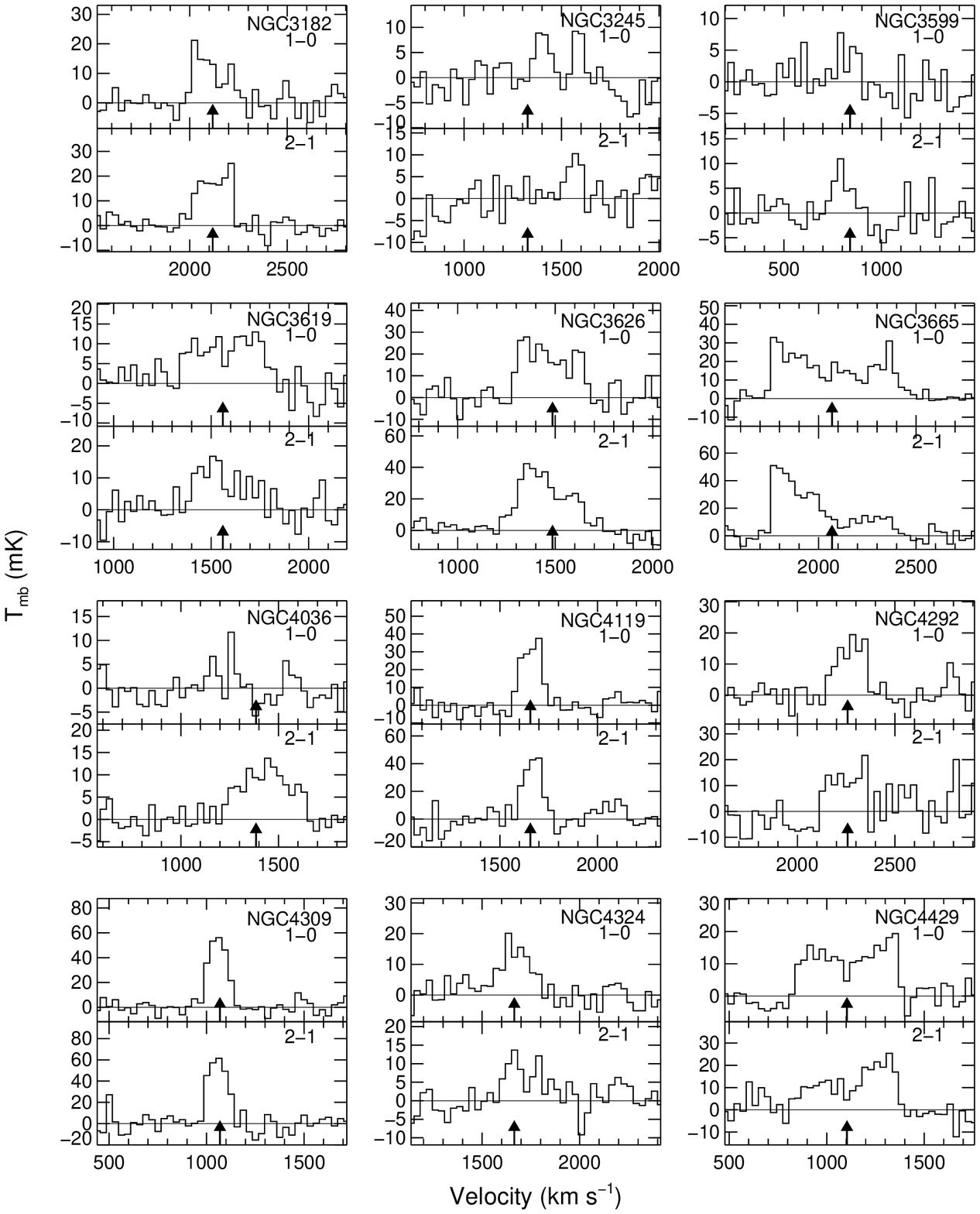}
\contcaption{}
\label{fig:spectra2}
\end{center}
\end{figure*}

\begin{figure*}
\begin{center}
\includegraphics[width=16cm,trim=1cm 0.7cm 1cm 1cm]{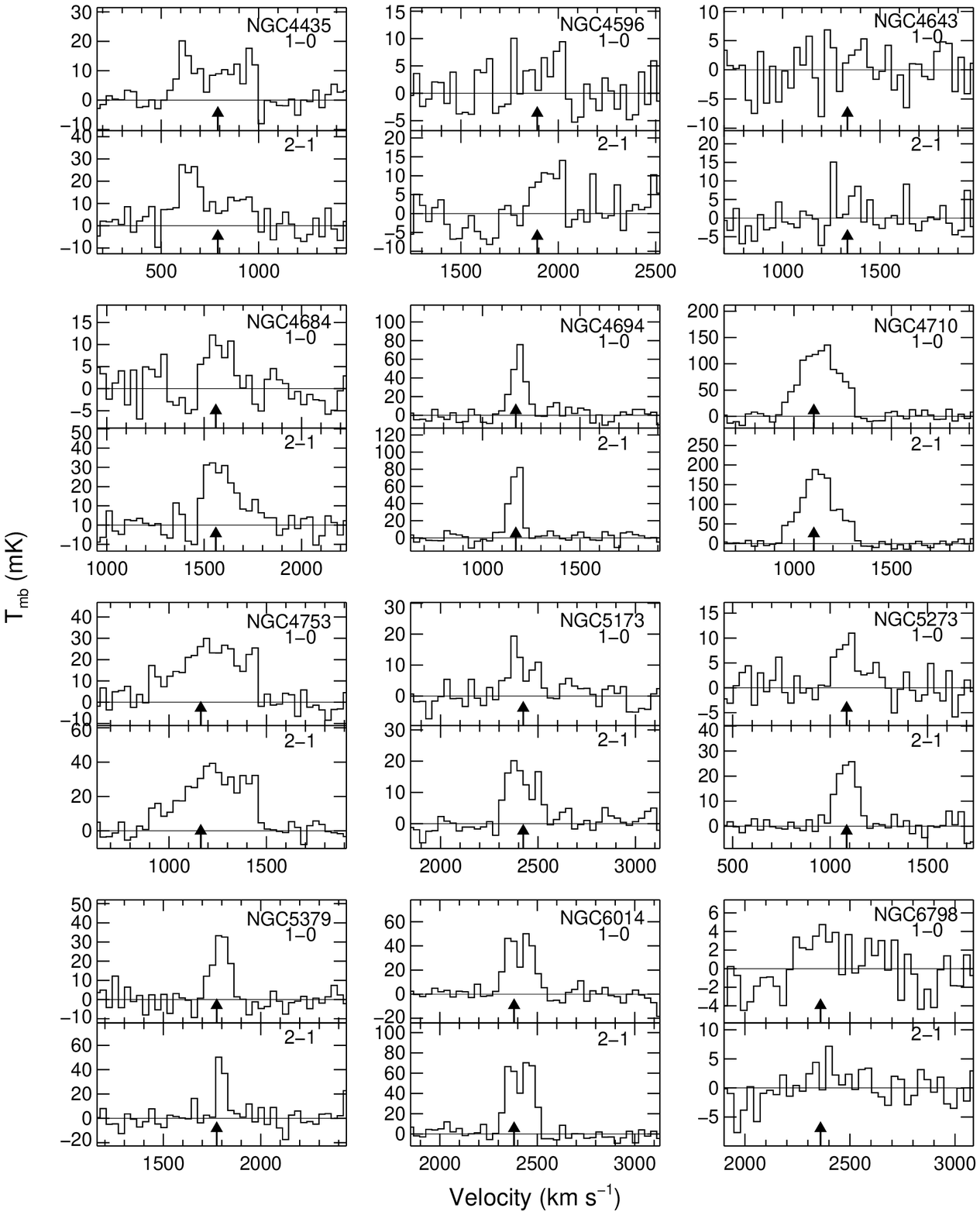}
\contcaption{}
\label{fig:spectra3}
\end{center}
\end{figure*}

\begin{figure*}
\begin{center}
\includegraphics[width=16cm,trim=1cm 0.7cm 1cm 1cm]{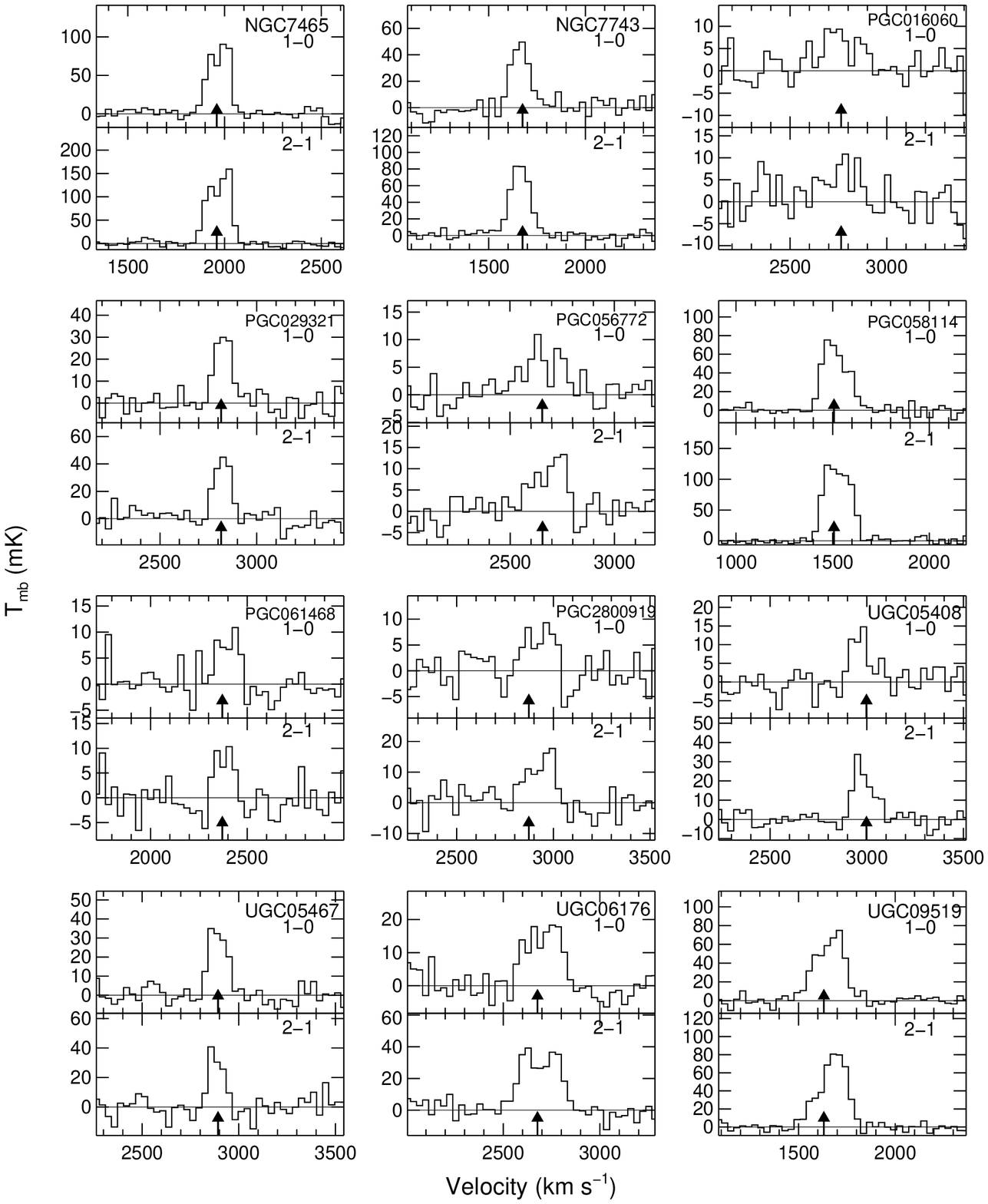}
\contcaption{}
\label{fig:spectra4}
\end{center}
\end{figure*}

Galaxies that were fit with a double-horned profile shape include NGC\,2764,
NGC\,3182, NGC\,3665, NGC\,4429, NGC\,4435, NGC\,4526, and UGC\,06176.  Other galaxies
that show signs of a double-horned shape such as a flat top and shoulders above
the Gaussian, or steeper sides than the Gaussian, include IC\,1024, NGC\,524, NGC\,3626,
NGC\,4119, NGC\,4281, NGC\,4596, NGC\,4643, NGC\,6014, NGC\,7465, PGC\,058114, and
UGC\,05408.
The classic mechanism for producing a double-horned shape is that in a disc with a
flat rotation curve, the material along the major axis
(near the line of nodes) all has much the same projected velocity and it is the
extreme velocity \citep{wiklind97}. 
Thus we find that 
almost a third of the detections show signs that the gas is in a regular disc
with a flat rotation curve.
Profiles that do not show a double-horned shape could indicate that the gas does
not extend as far as the flat part of the rotation curve, that the disc is
face-on, the gas is not in a disc, or that the signal-to-noise ratio of the line is simply too low.
Thus, the true incidence of regular discs is likely
to be larger than 30\%.

At least 13 of the detected lines are visibly asymmetric, and again that must be a
lower limit since the detectability of asymmetries requires relatively high
signal-to-noise ratios.  These asymmetries could arise from a gas
distribution which is intrinsically lopsided (as if the gas were recently acquired
and has not yet settled into dynamic equilibrium).  They could also arise if
the gas's spatial extent is comparable to or larger than the size of the beam and pointing errors
cause some of the gas, even in a regular disc, to fall outside of the beam.
Notable examples of asymmetric profiles
include IC\,1024, NGC\,1222, NGC\,3626, NGC\,3665, NGC\,4335, NGC\,4573, NGC\,4694,
NGC\,4710, NGC\,5173,
NGC\,7465, PGC\,056772, PGC\,058114, UGC\,05408, and UGC\,09519.
NGC\,3607 and NGC\,5866 \citep{WS03} also fall into this category, and their gas is known to be
spatially extended from detections in multiple pointings.

Systemic velocites for the CO lines (Table \ref{tab:detections}) are in good
agreement with our newly measured optical velocities; the difference between those
two velocities has a dispersion of only 16 \kms.  Outliers, for which the velocity
difference is greater than 50 \kms, include only cases in which our central
pointing is not expected to have recovered all of the CO emission (NGC\,2685)
or cases in which the CO line is faint, so that it might be
intrinsically asymmetric or double-horned (NGC\,6798, NGC\,3245).  If there were
members of the sample that had recently acquired their gas from an external
source, they might have revealed themselves via molecular gas which had not yet
settled into the center of the galaxy's potential.  However, we find no convincing
evidence for any of these cases.  More detailed probes of the dynamical status of
the molecular gas will require interferometric CO maps.

\begin{figure} % line ratios
\begin{center}
\includegraphics[width=8cm,trim=2cm 0.5cm 1cm 1cm]{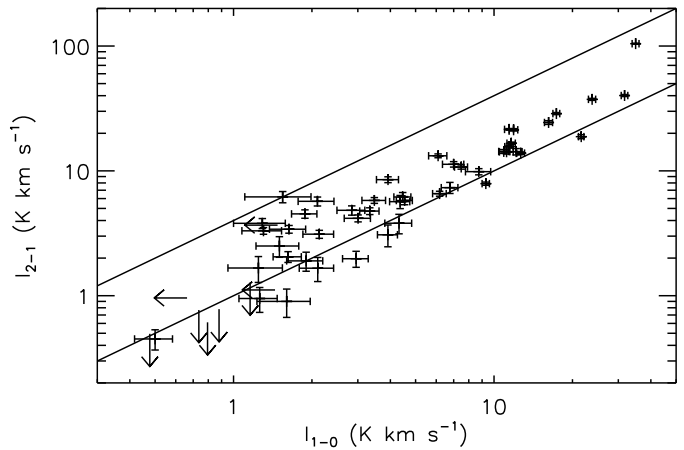}
\caption{Integrated intensities of CO(1-0) and CO(2-1) for the detected galaxies.  For
those detected in one transition but not the other, an arrow is plotted with the tail at
the $3\sigma$ formal limit in the non-detected transition.  Two indicative lines are also plotted,
the bottom
one showing $I_{1-0} = I_{2-1}$ and the top showing $I_{2-1} = 4I_{1-0}$.}
\label{fig:lineratio}
\end{center}
\end{figure}

The integrated CO(2-1)/CO(1-0) emission ratio is listed in
Table~\ref{tab:intensities}, and
Figure~\ref{fig:lineratio} plots the integrated CO(2-1) intensity against the
CO(1-0) intensity when at least one line is detected.  
Since only one central beam has been observed, the spatial region
covered is approximately four times smaller in CO(2-1) than in CO(1-0) and the line
ratio is affected by both the excitation temperature and the spatial distribution
of the gas.   
If the CO emission uniformly fills both the CO(1-0) and the CO(2-1) beam 
and both lines are optically thick, with the same excitation temperature,
one expects the integrated
intensity in K km/s to be the same in both lines.   This result occurs because, by
definition, the measured brightness
temperature is a specific intensity averaged over the beam area.  If, however, the CO emission is
compact compared to the beams (point spread functions) then the measured intensity in the CO(2-1) line should be
larger by up to a factor of 4.  
Subthermal excitation \citep[e.g.][]{brainecombes92} would decrease the 
intensity of the J=2-1 line relative to J=1-0, if densities are not large enough to
populate the upper rotational levels.

Most data points in Figure \ref{fig:lineratio} have $I_{2-1}/I_{1-0}$ between 1 and
4, suggesting that
in most cases the emission does not fill both beams uniformly nor is it extremely
compact.  Intermediate size scales, in which the extent of the emission is
comparable to or smaller than the beams, are probably the most common.
Two interesting outliers are NGC\,1266 and NGC\,4684, which have
quite high CO(2-1)/CO(1-0) line ratios ($\ge 3$),
suggesting compact gas distributions.
A few cases show the intensity of the CO(2-1) line to be anomalously low,
{\it i.e.} below the 1:1 relation.  These cases could be due to either subthermal
excitation or 
pointing errors which cause the intensity of the 2-1 line to be underestimated.

\begin{figure} % gas surface densities
\begin{center}
\includegraphics[width=8cm,trim=2cm 1cm 1cm 1cm]{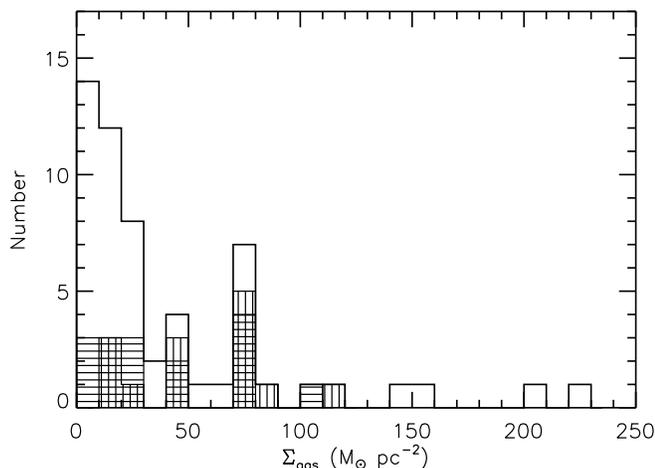}
\caption{Average molecular surface densities for the detected galaxies.  The open
histogram shows all detected galaxies.  Horizontal and vertical lines show galaxies
with double-horned and asymmetric profiles, respectively.}
\label{fig:siggas}
\end{center}
\end{figure}

Molecular gas in nearby spirals is observed to have surface densities ranging
from $> 10^3$~\msunsqpc\ in some galaxy nuclei to a few tens of \msunsqpc\ in
the outer discs \citep{regan06}.
Indeed, sensitive observations of both HI and CO emission in spirals suggest
that a typical cold gas disc in a spiral becomes dominated by molecular gas
when the total gas surface density is greater than about 14 \msunsqpc\
\citep{leroy08}.
For comparison, Figure \ref{fig:siggas} shows the distribution of the average
molecular surface densities among the CO detections of our sample.  Since our
data consist of a single pointing only, those surface
densities are averages over the 30m beam (2 to 4 kpc in diameter), based on the intensity of the 1-0
line.  Mass densities are multiplied by a factor of 1.36 to account for helium.
The molecular surface densities in the CO-rich \atlas\ galaxies are well
within the range occupied by nearby spirals, and 25\% of our detections have
average surface densities greater than 50 \msunsqpc.  Galaxies showing signs of
double-horned or asymmetric profiles (Table \ref{tab:detections}) have a surface
density distribution which resembles that of the whole sample, with the caveat that
such signs cannot be identified if the signal-to-noise ratio is low.  The apparent
dearth of double-horned or asymmetric profiles at small surface densities may only be
an effect of the noise level.
These considerations
suggest that the CO discs in early-type galaxies probably have gross
characteristics which are not too different from the molecular discs in
spirals.

The galaxies with the highest average surface densities are NGC\,1266 (230
\msunsqpc) and NGC\,4710
(210 \msunsqpc).  Since the line ratios for NGC\,1266 already indicate that the
gas distribution is more compact than the 1-0 beam, this galaxy's true molecular
surface density is even higher than 230 \msunsqpc.  In fact, high
resolution interferometric observations show the molecular gas in the galaxy is so
compact that the surface densities are on the order of $10^4$ \msunsqpc\
\citep{katey1266}.

\subsection{CO linewidths}

As mentioned above, the CO line shapes and widths carry 
information about the gas kinematics and radial extent, and appropriate analysis
of the linewidth distribution can help extract this information at least in a
statistical sense.   
The 1300 \kms\ bandwidth should be large enough to detect all reasonable CO
lines, if the signal-to-noise ratio were not a limitation.  In practice, of course, 
for a fixed \htoo\ mass, spreading the emission over a larger velocity range will
decrease the amplitude of the peak and make the line more difficult to
detect.
And since a broad CO line implies a massive host galaxy with a deep potential well,
a bias against the detection of broad lines could potentially affect the
statistical analysis of molecular mass distributions.  

In this context it is useful to consider the effects of our analysis procedures on
the detectability of faint emission lines.  A line of low amplitude may erroneously
be included in the estimate of the spectral baseline level.  Then, in searching for
detections, we sum the baseline-subtracted spectrum over a 300 \kms\ velocity
range, which (of course) will not be the entire line if the linewidth is larger
than 300 \kms.  Both of these procedures make the recovered line area
systematically smaller than the ``true" line area for weak lines.  However, for
lines of 300 \kms\ or narrower, the recovered area is a constant fraction of the
``true" area regardless of the linewidth.  Monte Carlo simulations of our analysis
procedure confirm this expected result.  The simulations also confirm the strong
bias against broader lines (at a fixed line area)
for linewidths greater than 300 \kms.  For example, typical nondetections in our
sample have thermal noise levels such that an integral over 300 \kms\ has an uncertainty
of 0.33 \kkms.  A line of 3.3 \kkms\ area and 600 \kms\ width (amplitude 5.5 mK),
and a typical thermal noise level, actually has only a 40\% chance of being
recovered as a detection by our analysis procedures if it is mistakenly included in
the baseline estimate, though it has a 97\% chance if it isn't included in the baseline.
Thus, for lines of 300 \kms\ or narrower
there should be no bias in the detectability as a function of linewidth (for a
fixed line area), though for lines broader than 300 \kms\ the effect can be severe
at small amplitudes.

\begin{figure} % linewidths
\begin{center}
\includegraphics[width=8cm,trim=1.5cm 0.7cm 1cm 0.7cm]{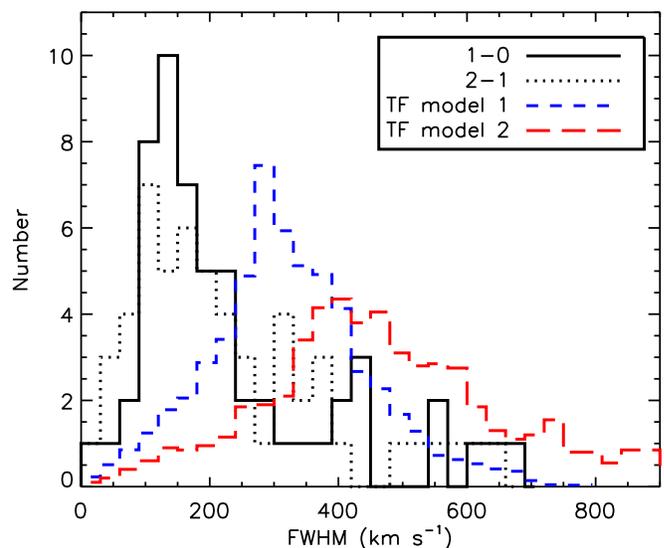}
\caption{Observed and modeled CO linewidths.  The models are described in the text.
}
\label{fig:linewidths}
\end{center}
\end{figure}

Figure~\ref{fig:linewidths} shows the observed distribution of CO linewidths in the
\atlas\ sample.
As described in section \ref{sec:distrib}, these are the FWHM of either a
fitted Gaussian or a double-horned profile (in fact both methods give the
same width for profiles which are obvious double-horns).
Three galaxies which each have two features detected at greater than 3$\sigma$
formal significance in a Gaussian fit are here interpreted as faint
double-horned profiles; they are NGC\,4281 and NGC\,4643 in J=2-1 (634 \kms\ and
366 \kms\ wide, respectively) and NGC\,4596 in J=1-0 (231 \kms\ wide).
Six detections from \citet{WS03} and \citet{SWY1} do not have fitted linewidths
quoted in their discovery papers, and in these cases the linewidths are estimated
from plots of the spectra; four of the six are asymmetric enough that fits would
be useless in any case.

The observed CO linewidths can be compared to simple model distributions generated as
follows.  In ``TF model 1," an estimated $V_c$ is generated for each \atlas\ 
galaxy from its \mk\ via the
S0 Tully-Fisher relation of \citet{williams2010}.
For each member of the sample we then generate 20 model CO linewidths as $2 V_c
\sin(i),$ where the inclinations are random and the probability of an inclination
$i$ is $p(i) = \sin(i)$.  The model CO linewidth distribution is scaled down in
amplitude by
the factor of 20 and by the global detection rate 0.22 to appear as the 
histogram in Figure~\ref{fig:linewidths}.  
In ``TF model 2," a similar procedure is followed except that $V_c$ is generated
from the galaxy's dynamical mass, again using the relations of
\citet{williams2010}, and in this case only the CO detections are used.
Additional details on the dynamical masses are given in Section \ref{sec:mdyn}.

These models show the expected CO
linewidth distributions under some non-trivial assumptions:
(1) that the Tully-Fisher
relations quoted above are appropriate; 
(2) that the CO is
always in relaxed discs which reach to the flat, asymptotic value of the circular
velocity curve; and
(3) in the case of model 1, that the CO-rich galaxies
have the same \mk\ and circular velocity distribution as the entire sample. 
The third assumption is addressed in Section \ref{sec:correl} below,
where we show that the CO detection rate is not dependent on \mk.  
There is a modest bias in the CO detection rate with dynamical mass, but model 2
already corrects for that bias by using only the CO detections.
Some justification for the second assumption
is suggested by the preponderance of double-horn profiles, but 
interferometric maps and independently-derived circular velocity profiles
will be required in order to check for which galaxies the assumption is not
satisfied.  Four cases are discussed in detail by \citet{YBC}.

It is apparent that we have detected many more relatively narrow lines, of FWHM 
100-200 \kms, than would be expected in either of the simple models.
Based on the above analysis of our detection procedures, we believe that the
observed peak at around 150 \kms\ 
and the associated drop in the histogram between 150 \kms\ and 300 \kms\
are not imposed by our methodology.
They could be related to a breakdown of the assumptions, which would
mean that we don't have an accurate estimate of $V_c$ or the CO doesn't extend far
enough in radius to trace $V_c$ in some cases.
Using the profile full width at 20\% of the peak, rather than at 50\%, would
eliminate some of the discrepancy but not all of it.  For Gaussian profiles that
change would increase the measured width by a factor of 1.5, but for double-horn
profiles it should not affect the measured width.
There are also several observational effects which would tend to produce an 
overabundance of narrow lines.
Narrow lines will arise if the CO is extended and pointing errors prevent the
detection of one horn of a double-horned profile (as in an extreme version of our
CO 2-1 spectrum of NGC\,3665).  If the CO is in a ring whose radius is larger than
the 30m beam, a narrow line may also result.  Based on optical images, this effect 
may be a problem for a galaxy such as NGC\,5379.  Both of these discrepancies would
also be rectified with interferometric maps.  
One impact of these effects is that
care is required when using the CO linewidths for a Tully-Fisher study
\citep{TimTF}.

\section{MOLECULAR MASS CORRELATIONS}
\label{sec:correl}

One of the major unsolved questions concerning the molecular gas in 
early-type galaxies is its origin -- whether it has been present in
one form or another since the galaxies were assembled or whether it was  more
recently acquired from some external source.
The \atlas\ data give a wealth of information on the structural, dynamical, and
stellar population properties
of the galaxies, or in other words, their assembly
and star formation histories.  Examination of which types of
galaxies are gas-rich and which are gas-poor may help to address the origins of
the gas and, by implication, its role in the evolution of different kinds of
early-type galaxies.  Figures~\ref{fig:correlations} through \ref{fig:xray}
compare the molecular gas contents to various other properties of the host galaxies.
We discuss the implications of these results below.

\subsection{Stellar luminosity}

\begin{figure*} % first red dot plot
\begin{center}
\includegraphics[width=16cm,trim=2cm 1cm 1cm 1cm]{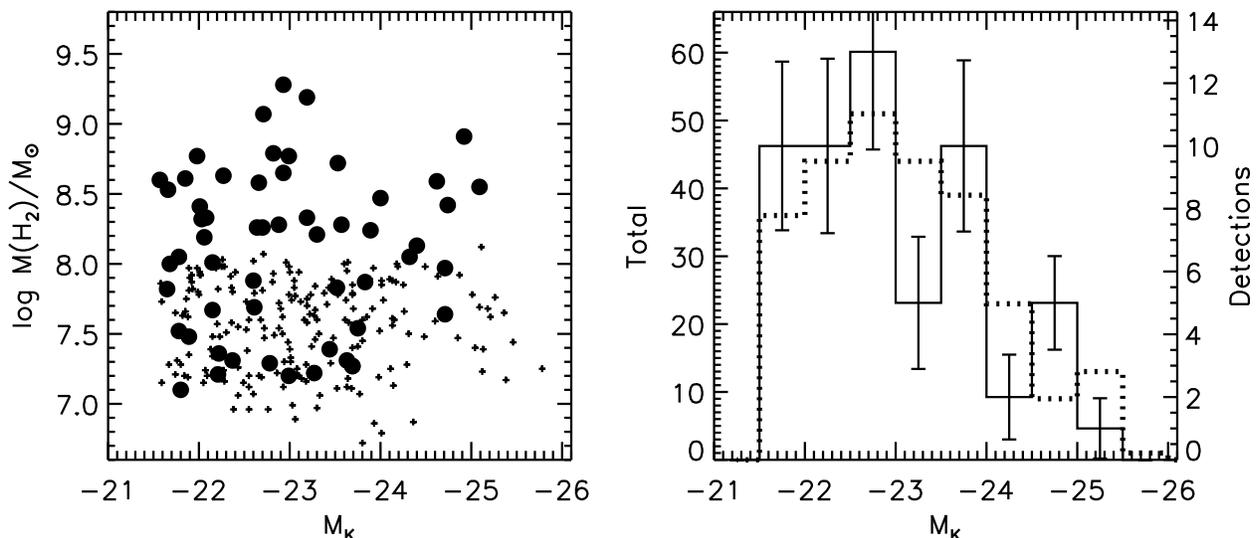}
\caption{Molecular masses and \mk.
Small crosses are
\atlas\ sample galaxies that are not detected in CO emission (3$\sigma$ upper limits), 
and large circles are detections.
The histograms in the right-hand panel show both the
properties of the entire \atlas\ sample (dotted line; left-side scale) and those
of the galaxies detected in CO (solid line; right-side scale).  The relative
scaling for the two histograms is the global detection rate.
Binomial uncertainties are indicated for the histogram of detections.
\label{fig:correlations} }
\end{center}
\end{figure*}

Figure \ref{fig:correlations} shows the distribution of molecular mass with $K$-band
luminosity and a histogram of \mk\ for the CO-detected galaxies.
CO emission is detected over nearly the entire range of
luminosities in the sample, and there is no clear trend in \mhtoo\ versus \mk.
There is a small hint of a decrease in the detection rate at the highest
luminosities.  Only one of the 13 galaxies with $M_K < -25$ 
is detected, whereas 3 would have been expected at the global average
detection rate.  However, the statistical significance of this difference is low.  
A stellar mass of $10^{11}$~\solmass\ corresponds to $M_K = -24.4,$ at a typical
stellar mass-to-light ratio of $M_*/L_K = 0.82$ \citep{bell03}; the CO detection
rate among galaxies brighter than this is 6/25 = 0.24 \error 0.09, consistent with the
global average.

A Kolmogorov--Smirnov (KS) test on the $M_K$ distributions of 
CO detections and nondetections indicates that they are consistent with each other
(the probability that those two $M_K$ distributions could have been drawn from the
same parent population is 0.52).
Thus we infer a constant CO detection 
rate at all luminosities from $M_K = -21.5$ to $-26$.  Refuting that model
will require observations of many more galaxies brighter than $M_K =
-25$, but such galaxies are rare in the local universe.  

\begin{figure} % km on luminosities
\begin{center}
\includegraphics[width=8cm,trim=1.5cm 0.7cm 1cm 1cm]{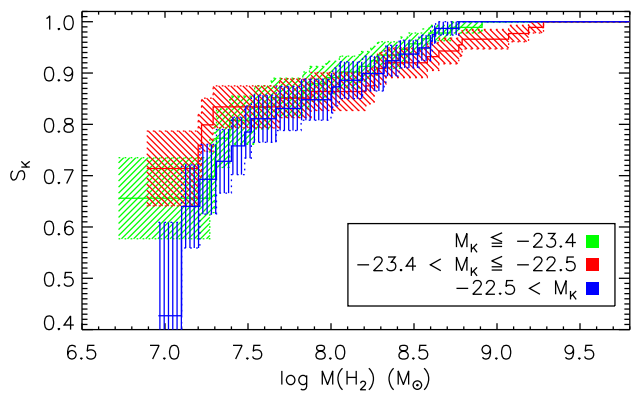}
\caption{The Kaplan-Meier estimators for the cumulative \mhtoo\ distribution
functions $(S_K)$ for high-, medium-, and low-luminosity \atlas\ galaxies.  These
estimators include the effects of censored data (upper limits) in constructing 
the cumulative distribution function.  The
shaded regions indicate $1\sigma$ uncertainties on the estimated cumulative distribution
functions.  The distribution functions are similar for all three luminosity
ranges, suggesting no dependence of molecular mass on host luminosity.}
\label{km2}
\end{center}
\end{figure}

The dependence of molecular mass on galaxy luminosity can also be investigated by
comparing the cumulative \htoo\ distribution functions for high, medium,
and low luminosity galaxies (Figure~\ref{km2}).  Since the bulk of the CO observations produced
non-detections (``censored" data) , the appropriate statistical tool is the Kaplan-Meier estimator for
the cumulative distribution function of a randomly censored sample.  We calculate
the Kaplan-Meier estimator using the software packages
ASURV version 1.3 \citep{ASURV,FN85} and R \citep{Rcite}, with consistent answers
from both packages.  Here we
divide the \atlas\ sample approximately into thirds by luminosity, i.e.\ $M_K
\leq -23.4,$ $-23.4 < M_K \leq -22.5,$ and $-22.5 < M_K.$ 
The CO detection rate in these bins is 19/92, 17/88, and 20/79, respectively.
Figure~\ref{km2} shows that the \mhtoo\ distributions for these three luminosity
bins are all consistent with each other, an impression that is confirmed by the
Gehan, logrank, and Peto two-sample tests.  Specifically, in the assumption that
the \htoo\ mass distributions for the bright and faint galaxies are drawn from the
same underlying population, the probability of measuring a difference as large as 
the one observed is estimated at 0.57 to 0.60.
For comparison, the ``dichotomy" stellar mass of $3\times 10^{10}$ \solmass\
\citep{kauffmann03} occurs at $M_K \sim -23.1,$ for $M_*/L_K = 0.82$
\citep{bell03}; this stellar mass is in the middle luminosity bin.
Our analysis 
indicates no measurable difference in the \htoo\ content of galaxies as a
function of their stellar luminosity.

The use of the Kaplan-Meier methodology merits some discussion, as it assumes
the censoring pattern is random \citep[e.g.][]{walljenkins}.   In this case the censoring in brightness
temperature is certainly not random, as observations were made to a fixed brightness
temperature noise level.  However the censoring in \htoo\ mass and \mhtoolk\ should
be approximately random, for the reasons discussed by \citet{if92},
\citet{osullivan}, and \citet{walljenkins}.  Multiplication by the
square of distance and division by the optical luminosity effectively serve to 
randomize the censoring pattern in those derived quantities, since our sample spans
a range of distances and there is no hint of a dependence of CO emission
on optical luminosity in our sample (Figure \ref{fig:correlations}).  Thus we believe the assumption of random censoring
is justified.  In addition, our objectives are to assess the
similarity of the cumulative distribution functions of different subsets of our data
and there is no reason to believe that the censoring should differ from one subset
to another.

As mentioned above, the practical difficulty of detecting a broad and weak
spectral line means that a given mass of molecular gas will be easier to detect
in a face-on galaxy or in an edge-on galaxy with a small circular velocity than in an edge-on galaxy
with a circular velocity of 200 \kms\ or higher.  This effect could cause the
\htoo\ mass distribution to be underestimated in the highest luminosity bin.
Careful simulations would be required to gauge the magnitude of the effect, but
the sense of the correction would be to increase the incidence of molecular gas in
the highest luminosity galaxies.

The lack of a correlation between \htoo\ mass and stellar luminosity has 
been interpreted in the past to mean that the molecular gas is unrelated to internal
stellar mass loss \citep[e.g.][]{kr96,wch95}.  
The assumption is that internal mass loss would produce
a linear relationship between stellar luminosity and \htoo\ mass.
However, if the mass loss material is shock-heated to $10^6$~K or higher by the
relative stellar velocities or is otherwise heated by ambient hot gas,
it would be difficult to predict how much of it might
be able to cool and re-form molecules.  Environmental and feedback effects would
also complicate the relationship between the stellar luminosity and the amount of
molecular gas present in a galaxy.  Thus, we argue that one cannot infer either
internal or external origin for the gas based on the \mhtoo-\mk\ relationship 
alone (or the lack of one). The value of the \atlas\ project is, of course, that other
structural and kinematic
parameters and stellar population information are available for the sample as
well.

\subsection{Dynamical mass}\label{sec:mdyn}

\begin{figure*} % second red dot plot
\begin{center}
\includegraphics[width=16cm,trim=2cm 1cm 1cm 1cm]{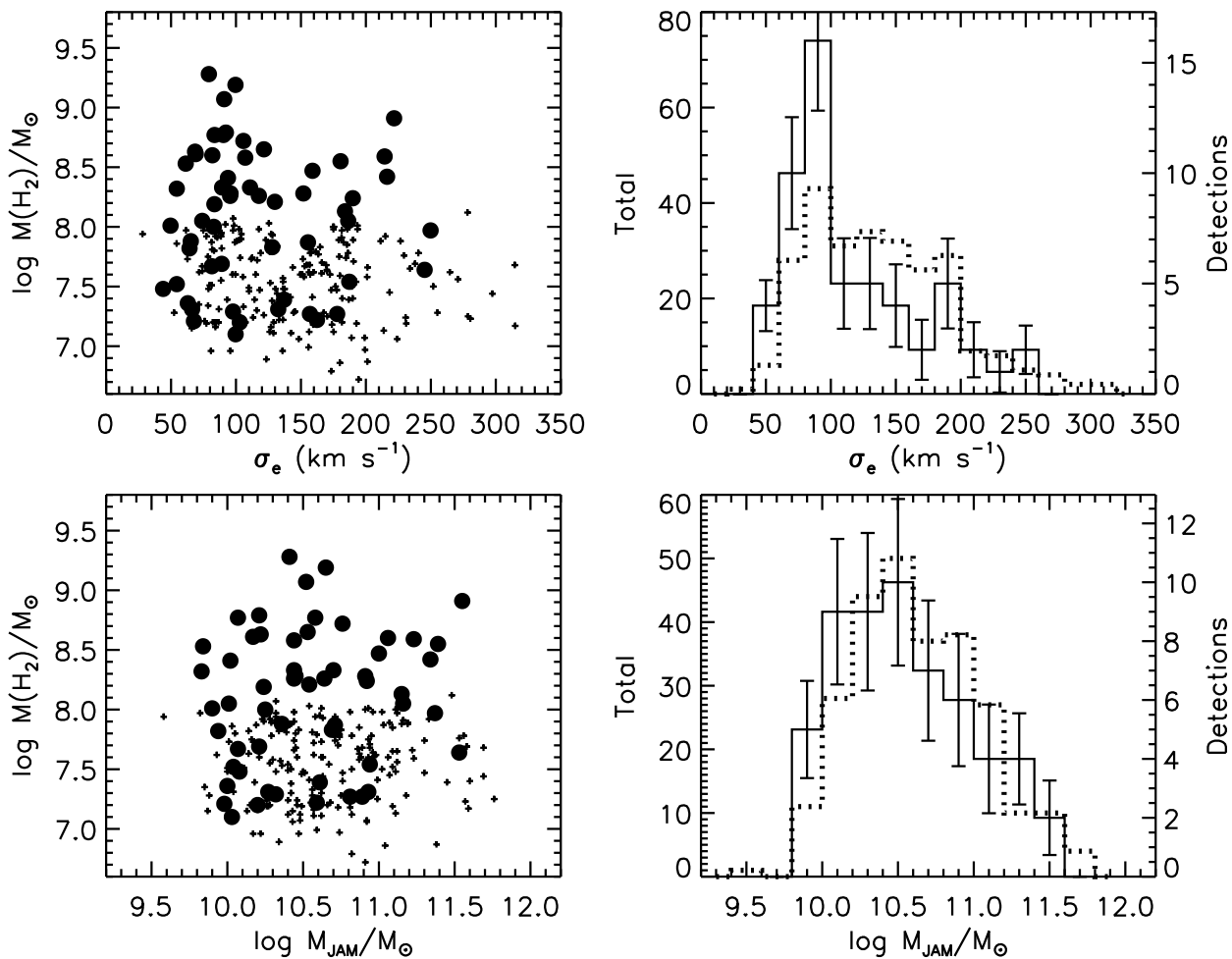}
\caption{Molecular mass, dynamical mass, and velocity width \sigmae.  Symbols and lines are as in
Figure~\ref{fig:correlations}.  Dynamical masses are obtained from Jeans
anisotropic models (Cappellari et al.\ in prep) as described in \citet{JAMmethod}.
\label{fig:mdyn} }
\end{center}
\end{figure*}

In spite of the lack of a relationship between CO content and \mk,
there is some evidence for a modest dependence of the CO content on the dynamical mass and a closely
related quantity, the global stellar velocity width \sigmae.
Figure~\ref{fig:mdyn} shows that the CO detection rate for galaxies with
$\sigma_e \le 100$~\kms\ is relatively high, 0.38 \error\ 0.05, whereas for $\sigma_e >
100$~\kms\ it is 0.14 \error\ 0.03.
The median \sigmae\ for CO detections is 99 \kms, whereas for nondetections it is 137
\kms.  The effect in dynamical mass is more subtle, so that the CO detection rate
above a dynamical mass of $3\times 10^{10}$~\solmass\ is 0.19 \error\ 0.03 and below
is 0.25 \error\ 0.04.  A KS test on the dynamical masses of CO detections and
nondetections gives a probability of 6.5\% that they could have been drawn from the
same parent population.  However, the Kaplan-Meier estimators show no significant
differences in \htoo\ masses of high and low mass galaxies.

For both \sigmae\ and dynamical mass the sense of the difference is that lower mass
galaxies are more likely to be detected.
It is curious that the magnitude of the effect is much stronger in
\sigmae\ than in dynamical mass.  The details of the scaling relations between
those two quantities are beyond the scope of this paper, but it is helpful to
be precise about their definitions.
\citet{Spaper4} explain that \sigmae\ is the second moment of the galaxy's
luminosity-weighted line of sight velocity distribution (LOSVD).  Because it is
constructed by fitting a Gaussian to a stack of all spectra within the effective
radius, it includes contributions from both the mean stellar rotation and the
velocity dispersion approximately as $\sigma_e \sim (V_{rot}^2 + \sigma^2)^{1/2},$
where $V_{rot}$ is the projected mean stellar rotation velocity and $\sigma$ is
the local stellar velocity dispersion.  \citet{Spaper4} discusses the use of
\sigmae\ in the virial mass as $M_{vir} \sim 5\sigma_e^2 R_e/G.$ 
The dynamical mass used here is, however, derived from stellar dynamical models 
of the SAURON stellar kinematics \citep{MCHiA}, using the Jeans anisotropic
modelling technique of \citet{JAMmethod}.
It can be understood as 
$\mjam \approx 2\times M_{1/2},$ where $M_{1/2}$ is the total dynamical mass
within a sphere containing half of the galaxy light.  In addition, as $M_{1/2}$ is
generally dominated by the stellar mass, $\mjam$ approximates the stellar mass
$M_{star}$ but should not be conflated with the total halo mass out to the virial
radius.  The quantities $\mjam$ and
\sigmae\ for the \atlas\ sample will be published in Cappellari et al., in
preparation.

Several different effects could be responsible for the dependence of the CO
detection rate on \sigmae\ and dynamical mass.  
We note first that molecular gas likely does induce
star formation, and the younger stellar populations should be both brighter and
dynamically colder than the rest of the stars.  Star formation should {\it not} affect
the dynamical mass or
\sigmae, though, as \sigmae\ already includes both the rotation and the dispersion components of
the velocity width.  Instead, the dependence of the CO detection rate on dynamical
mass and \sigmae\
could be driven by either or both of (1) an observational bias against the detection
of broad CO lines, as described above, and (2) a downsizing or rejuvenation effect in which the more
massive galaxies are less likely to contain molecular gas.
The observational bias would manifest itself as a
deficit of CO detections in edge-on high mass galaxies compared to
face-on high mass galaxies.  A downsizing effect would be consistent with much other
recent work on star formation histories as a function of galaxy mass
\citep[e.g.][]{rogers10,zhu10,thomas10}.  

Of course, regardless of whether the dependence on dynamical mass
is an observational bias or a real downsizing effect (or both), something of a
`conspiracy' between the mass and the mass-to-light ratio is required in order to
reproduce the result that the CO detection rate is completely independent of \mk.
This issue can also be investigated in greater detail with estimates of the stellar
population mass-to-light ratios of the \atlas\ galaxies. 

\subsection{Morphological type}

There is a significant difference in the CO detection rate between the galaxies
traditionally classified as elliptical and lenticular.
The \atlas\ sample contains 68 galaxies whose morphological type in LEDA is $T <
-3.5$ (ellipticals).  
Only 3 of the 56 total CO detections are in this group, for a detection rate of
0.05 \error\ 0.03.  These detections are NGC\,2768,
NGC\,5173, and NGC\,4283, and the latter is of marginal quality
\citep{SWY1}.
However, we caution that these morphological types are generally based only on 
photographic imaging and have the limitations (or perhaps ``features") that they
carry little kinematic information and that 
the presence of a dust lane 
in the optical photographs could cause a galaxy to be classified as lenticular 
rather than elliptical.

Interferometric imaging shows that the molecular gas in early-type galaxies is
invariably associated with dust lanes \citep{y02, y05, YBC}, so that if the
morphological classification is indeed biased by the presence of a dust lane then
it is nearly a foregone conclusion to find a higher CO detection rate in
lenticulars than in ellipticals.
Indeed, Paper~II 
tabulates the evidence for dust discs, dusty filaments, and blue regions in optical
images of the \atlas\ galaxies.   Of the galaxies with one or more of such features,
3/49 (6\%) are classified as elliptical and 46/49 as lenticular; but of the galaxies
with no dust or blue features, 64/209 (31\%) are elliptical and 145/209 are
lenticular. 
Of course these statistics do not prove a dust- or star formation-bias in
morphological classification, but they are consistent with such an interpretation.

From a different point of view, the SAURON observations of kinematics in early-type galaxies have
shown that there is often little if any structural difference between
lenticulars and ellipticals \citep{emsellem04}, so a purely isophotal 
classification contributes little to our understanding of their formation
histories.

\subsection{Specific angular momentum}

\citet{emsellem07} advocate that a more fundamental way to classify early-type
galaxies is with the $\lr$ parameter, a simplified and dimensionless version of the
luminosity weighted specific angular momentum.
It encapsulates some of the information on the degree of ordered vs.\ random
motions (i.e.\ rotational vs.\ pressure support) that one gets from the classic
$V_{rot}/\sigma$ ratio.
In addition, as some systems show rotation only 
within a few hundred pc of the nucleus, the $\lr$ parameter also reveals
something about the spatial extent of the rotation through the radial contribution
to the specific angular momentum.  Slow rotators (including galaxies with rotation
only at small radii) have small values of $\lr$.  In an update of this earlier
work, Paper~III 
demarcates fast and slow rotators at $\lr =
0.31\sqrt{\epsilon},$ where $\epsilon$ is the photometric ellipticity and both
quantities are measured as luminosity-weighted averages over the effective radius.
Paper~III also discusses the relationship between the traditional
E/S0 and the slow/fast rotator classifications.  There is some degree of alignment
between the systems, as the slow rotators are mostly classified as ellipticals.
However, the converse is not true, as 66\% of the ellipticals in \atlas\ are fast
rotators.  The consistency between the systems is only strong at the extremes of
$\lr$ and $\epsilon$.

\begin{figure*} % lambda red dot plot
\begin{center}
\includegraphics[width=16cm,trim=1.7cm 0.5cm 1cm 0cm]{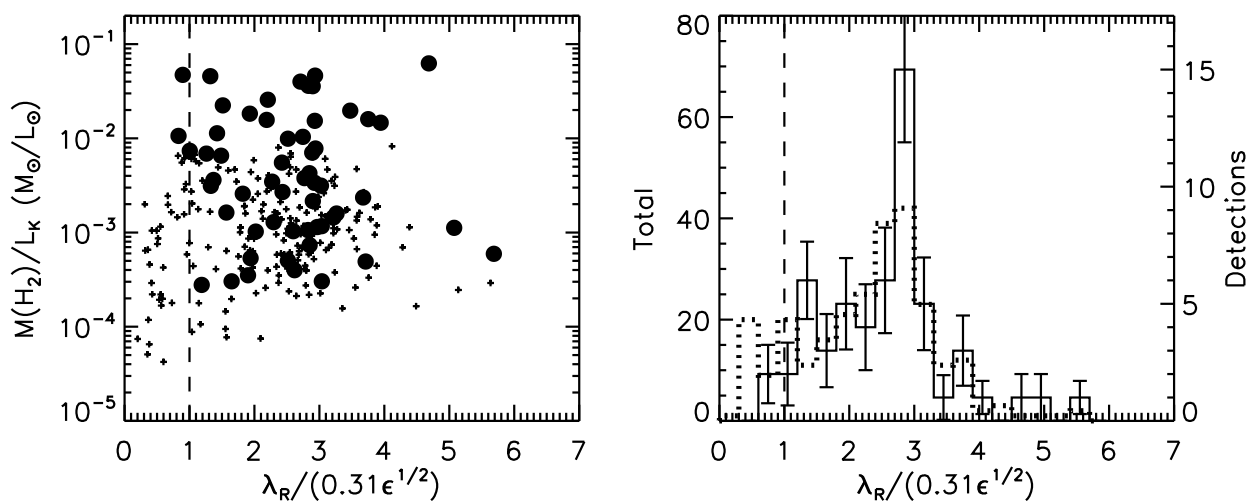}
\caption{\htoo\ and specific angular momentum.  The dashed line in both panels
separates fast and slow rotators (Paper III).  Other symbols and lines are as in
Figure~\ref{fig:correlations}.
\label{fig:lambda} }
\end{center}
\end{figure*}

Figure \ref{fig:lambda} shows normalized \htoo\ masses and CO detection rates as a
function of the ratio $\lreps$, and   
Table \ref{tab:classes} shows that
the CO detection rate among slow rotators (0.06\error 0.04) is significantly 
lower than that among fast rotators (0.24\error 0.03).  
The CO detections in slow rotators are NGC\,1222 and NGC\,4476.
Of these two, NGC\,1222 is kinematically disturbed due to a recent
interaction (Paper II).  NGC\,4476 also presents a complex structure as its
stellar velocities drop towards the outer
regions of the SAURON field and show signs of reversing at yet larger radii
(A.\ Crocker, private communication).
In short, there are few detections among the slow rotator class.  

Both a galaxy's value of $\lr$ and its projected circular velocity $V_c
\sin{i}$ are affected by inclination, of course.  However, 
{\it if} a galaxy has small $\lr$ due to being face-on, and if it has molecular
gas, it will have a relatively
narrow CO line and will be easy to detect.  Thus the deficit of CO
detections among galaxies of small $\lr$ is not due to inclination effects
because they have the wrong sense to reproduce such a deficit.  

The deficit of CO detections among galaxies of small $\lr$ is also not driven by the
trend with \sigmae\ or the dynamical mass.  Paper III %\citet{eric} 
shows that the slow rotators
do tend to be more massive than fast rotators, and they have higher \sigmae\ values, but
the overlap between fast and slow rotators is sufficient that this effect can be
controlled.  Of the 36 slow rotators in the \atlas\ sample, we begin with the subset
of 31 which have $\sigma_e <$~260 \kms\ or the 34 which have 
$9.5 < \log \mjam /M_\odot < 11.65$.  These subsets have CO detection rates of 0.06
\error\ 0.04, consistent with Table \ref{tab:classes}.  We construct a matched sample of fast rotators
by selecting (at random) one fast rotator which has \sigmae\ within \error 12 \kms\
or $\mjam$ within 0.1 dex of each slow rotator.  By design, then, the
matched sample of fast rotators has a distribution of $\mjam$ or \sigmae\ which is
statistically indistinguishable from that of the slow rotators.  Repeating the
process $10^4$ times, we find that the CO detection rate in our matched samples of
fast rotators is 0.24 \error\ 0.07 when the matching is done by \sigmae\ or 0.36
\error\ 0.07 when matched by $\mjam$.  Here quoted uncertainties are dispersions
in the set of $10^4$ trials.  In short, the CO detection rate is still significantly
higher among fast rotators than among slow rotators, even after controlling for the
\sigmae\ and $\mjam$ distributions.

\subsection{Internal kinematic structure}

Paper~II %\citet{Davor} 
documents photometric evidence for internal substructures such as
bars, rings, shells, tidal tails, and anemic spiral structures in the \atlas\
sample.  We find no evidence for a difference in CO detection rate between
galaxies showing these types of features and galaxies without them (Table
\ref{tab:classes}).
Similarly, we measure the misalignment angle between the
stellar photometric and kinematic axes, as these misalignments can be indications
of non-axisymmetric components like bars.
A KS test reveals no significant difference in the
misalignment angles of CO detections and nondetections.

Paper~II % \citet{Davor} 
also uses the features in the stellar velocity maps to classify the
internal kinematic structure of galaxies.  Galaxies are primarily classified as
``regular rotators," or RR, when the velocity field resembles that of a thin,
inclined disc such as a spiral galaxy.  Others are 
``non-regular," or NRR.  The RR class has a nearly one-to-one overlap with the 
fast rotators as defined by ellipticity and $\lr$, so these are assumed to be the
same family of galaxies.  Table \ref{tab:classes} shows that the CO detection rate
is significantly higher in the RR class than in the NRR class, and the `matched
sample' technique described in the previous section again confirms this result 
after controlling for different \sigmae\ and dynamical mass distributions of the RR and
NRR classes.

In terms of kinematic structure, early-type galaxies are quite
heterogeneous (Paper II); they include galaxies with no observable rotation
at all (non-rotators like M87), galaxies with kinematically distinct or 
even counterrotating
cores, galaxies comprised of two counterrotating stellar discs, such as IC\,0719,
and (infrequently) face-on disc-like regular rotators.  Table \ref{tab:classes} shows no particularly strong evidence that the CO
detection rate differs among these subtypes.  The CO detections in NRR galaxies
are NGC\,1222, NGC\,3073, IC\,0719, and NGC\,7465, of which NGC\,1222 is
also a slow rotator
and is discussed in the previous section.
Curiously, the ``two $\sigma$ peak" galaxies made of two 
counterrotating discs are exceptions to the general rule that
the CO detection rate is higher at low $\sigma_e$.   
This type has
the lowest dispersions $\sigma_e$ of any of the kinematic classes. 
Their relatively low CO 
detection rate (compared to the low \sigmae\ galaxies in general)
should not be an effect of the observational bias against broad lines because, given
their small masses, they
should not have broad CO lines.  

Papers~II and III clearly show that early-type galaxies
have a variety of different assembly histories.  Some general patterns of assembly
are conducive to retaining a high specific angular momentum within an effective
radius whereas other patterns are not.
The patterns which develop low specific angular momentum can include
misaligned major mergers, or mergers in which the orbital angular momentum cancels the spin
angular momenta of the progenitors (Bois et al.\ in prep), or a large number of randomly
oriented minor mergers.
Our analysis suggests that galaxies with these types of assembly histories are
least likely to be detected in CO.   
These may be the cases in which the misaligned angular momenta of gaseous discs
coming into the merger caused the gas to drop to the center of the galaxy where
it was consumed or destroyed.  Alternatively, they may have lost their cold gas
prior to their assembly or they may be less likely to acquire cold gas after
assembly.  Future work on gas kinematics and star formation histories can help to
distinguish between these scenarios.

%% New Table : detection stats of various morph. subtypes (see Davor paper)
%
\begin{table}
\begin{centering}
\caption{CO detection rates for kinematic types \label{tab:classes}}
\begin{tabular}{l c c c c}
\hline
Type      &    CO     &      no CO     &    Total    & CO det. rate \\
\hline
Fast          &  54  & 169 & 223 & 0.24\error 0.03\\
Slow          &   2  &  34 &  36 & 0.06\error 0.04\\
\\
regular rotators &  51  & 162 & 213 & 0.24\error 0.03\\
non-regular          &   4  &  40 &  44 & 0.09\error 0.04\\
\\
Ring/bar/shell/tail  &  23  &  79  & 102 &  0.23\error 0.04\\
No features          &  31  & 122  & 153 &  0.20\error 0.03\\
\\
\multicolumn{5}{c}{Kinematic subtypes}\\
Non-rotator   &   1  &   6 &   7 & 0.14\error 0.13\\
KDC/CRC       &   1  &  18 &  19 & 0.05\error 0.05\\
2 $\sigma$ peaks& 2  &   9 &  11 & 0.18\error 0.11\\
NRR/no feature  &   1  &  11 &  12 & 0.08\error 0.08\\
\\
\multicolumn{5}{c}{RR subtypes}\\
RR/2max       &   11 &  25 &  36 & 0.30\error 0.08\\
RR/other      &   40 & 137 & 177 & 0.23\error 0.03\\
\hline
\end{tabular}
\end{centering}

\textit{Notes:} Kinematic classifications are taken from Papers II and III.
This table and the next exclude NGC\,4486A, for which we have no CO
data.
\end{table}

\subsection{Environment: Virgo Cluster}

The \atlas\ sample contains 58 galaxies that are within 3.5 Mpc of M87 and are
considered members of the Virgo Cluster.  Many of these galaxies
(including M87) have distance measurements from surface brightness fluctuations
\citep{tonrySBF, mei07, paper1}.  When a surface brightness distance is
available, it is used in the computation of a galaxy's distance to M87.
Where independent distance measurements are not available for known Virgo
Cluster members, a distance of 16.5 Mpc is used as that is the adopted mean distance
of subclusters A and B \citep{mei07}.  For comparison, those authors also quote
the surface brightness
fluctuation distance to M87 as 17.2 Mpc, the typical distance uncertainties
for their data are 0.7 Mpc, and the line-of-sight depth of the cluster is 0.6 Mpc
\error 0.1 Mpc (1$\sigma$).
We note also that there is no statistically significant difference in the 
\mk, dynamical mass, or \sigmae\ distributions of
cluster and non-cluster galaxies in \atlas, so there is no detectable mass
segregation in this sample.

Twelve of the 57 Virgo Cluster members with CO data
are detected, for a cluster detection rate of 0.21\error 0.05 or less than
1$\sigma$ lower than the detection rate for the sample as a whole.  However, since
the Virgo Cluster is relatively nearby and we have observed our sample to a fixed
noise level, it is more appropriate to compare the detection rate in
Virgo to that of galaxies at a similar distance.  The entire \atlas\ sample
contains 89 galaxies at $D \leq 20$ Mpc and 124 galaxies at $D \leq 24$ Mpc; thus,
within 24 Mpc, half
of the sample galaxies are Virgo Cluster members and half are not.
A restricted comparison of galaxies within 24 Mpc mitigates the effect of
distance, since the typical \htoo\ mass limit at 24 Mpc is only a factor of 2
higher than at 17 Mpc.
Table \ref{tab:virgo} compares the CO detection statistics
inside and outside of the Virgo Cluster for all galaxies within 24 Mpc,
and also for the subset of fast rotators within 24 Mpc. 
The detection rate among cluster members (0.21 \error 0.05)
is nominally lower than the detection
rate of non-members (0.29 \error 0.06) and the detection rate of the 
entire sample (0.22 \error 0.03).  However, the
difference is at only the 1$\sigma$ level in the combined uncertainties
so more detailed investigations
of the CO luminosity functions are appropriate. 

%% Table 4: Virgo detection stats
%
\begin{table}
\caption{CO detection rates for galaxies within 24 Mpc\label{tab:virgo}}
\begin{tabular}{ l c c c c}
\hline
              &  CO  &  no CO  &  Total &  CO det. rate \\
\hline           
Virgo Cluster &  12  & 45  & 57 & 0.21\error 0.05 \\
non-cluster   &  19  & 47  & 66 & 0.29\error 0.06 \\
Total         &  31  & 92  & 123 & 0.25\error 0.04 \\
\\
\multicolumn{5}{c}{Fast rotators only} \\
Virgo Cluster   &  11  & 37  & 48 & 0.23\error 0.06 \\
non-cluster     &  19  & 42  & 61 & 0.31\error 0.06 \\
Total           &  30  & 79  & 109 & 0.27\error 0.04 \\
\hline
\end{tabular}
\end{table}

\begin{figure*} % environment/virgo red dots
\begin{center}
\includegraphics[width=16cm,trim=2cm 1cm 1cm 1cm]{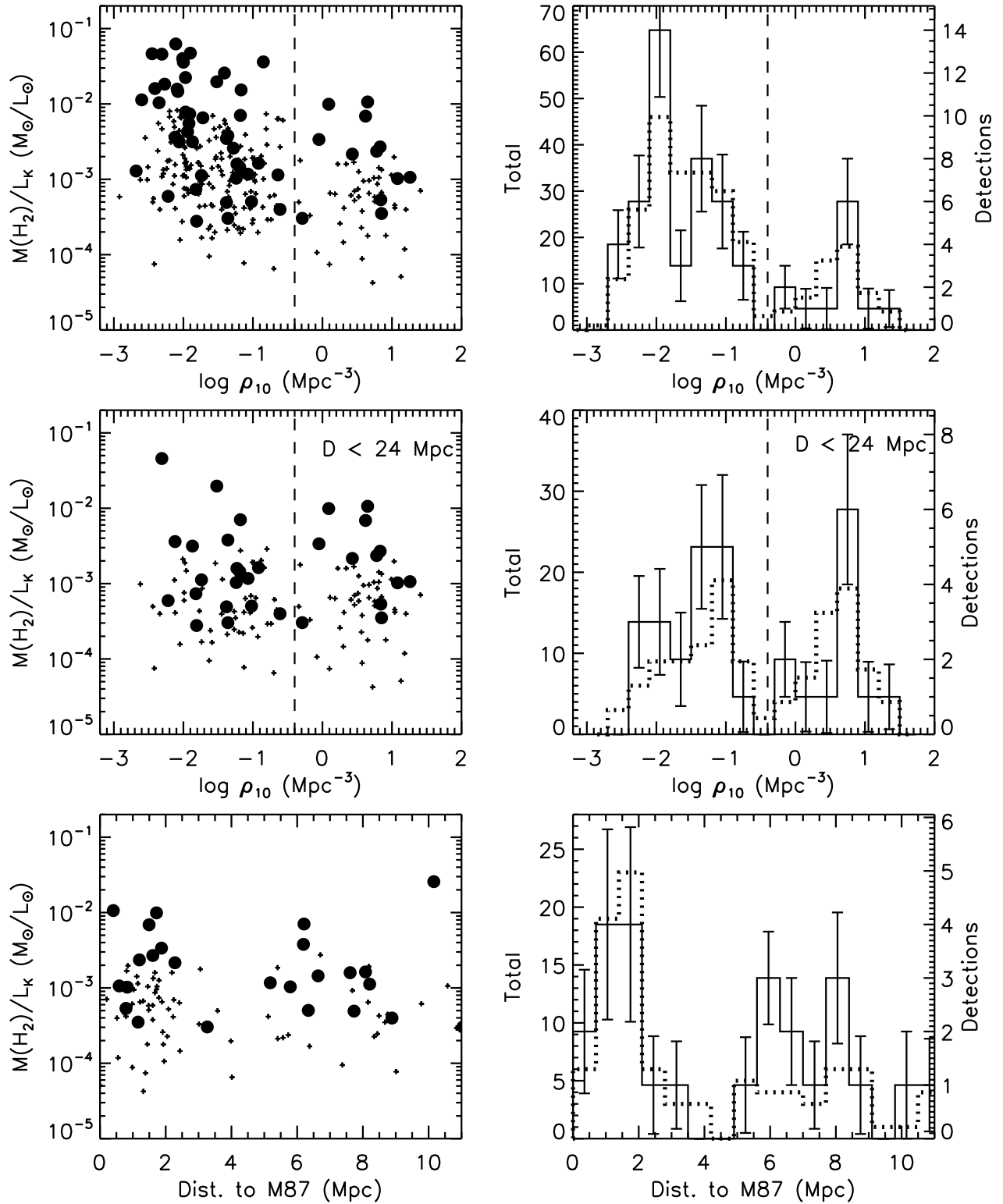}
\caption{Molecular gas properties inside and outside the Virgo Cluster.  The quantity
$\rho_{10}$ is a local galaxy density presented in Paper~VII, and the dashed line at
$\log \rho_{10} = -0.4$ separates Virgo Cluster members and nonmembers (Paper I).  Other symbols and
histograms are as in Figure~\ref{fig:correlations}.  The middle row is restricted to
galaxies with distances less than 24 Mpc (mitigating the effects of distance on CO
detection).  The third row presents molecular masses against the three-dimensional
(deprojected) distance to M87.
}
\label{fig:virgostats}
\end{center}
\end{figure*}

The top two rows of Figure~\ref{fig:virgostats} present the normalized
masses \mhtoolk\ versus the local bright galaxy density, both for the
entire \atlas\ sample and for the subset of galaxies closer than 24 Mpc.  Normalized
\htoo\ masses are also plotted as a function of the deprojected distance to M87.  Paper~VII
explains the calculation of the local density $\rho_{10}$ as the number density of
galaxies brighter than \mk\ = $-21.5$, averaged within a sphere which contains 10
such neighbors and quoted in units of Mpc$^{-3}$.  In the \atlas\ parent sample, which includes all morphological types,
all galaxies outside the Virgo Cluster have densities $\log \rho_{10} <
-0.4$ and all but two of the Virgo Cluster members have $\log \rho_{10} >
-0.4$.
A KS test on the distributions of local densities for CO
detections and nondetections (both within 24 Mpc; middle row, right side
panel of Figure~\ref{fig:virgostats}) gives a 31\% probability that they
could be drawn from the same parent population.

\begin{figure} % km in and out of virgo
\begin{center}
\includegraphics[width=8cm,trim=1.5cm 1.5cm 1cm 1cm]{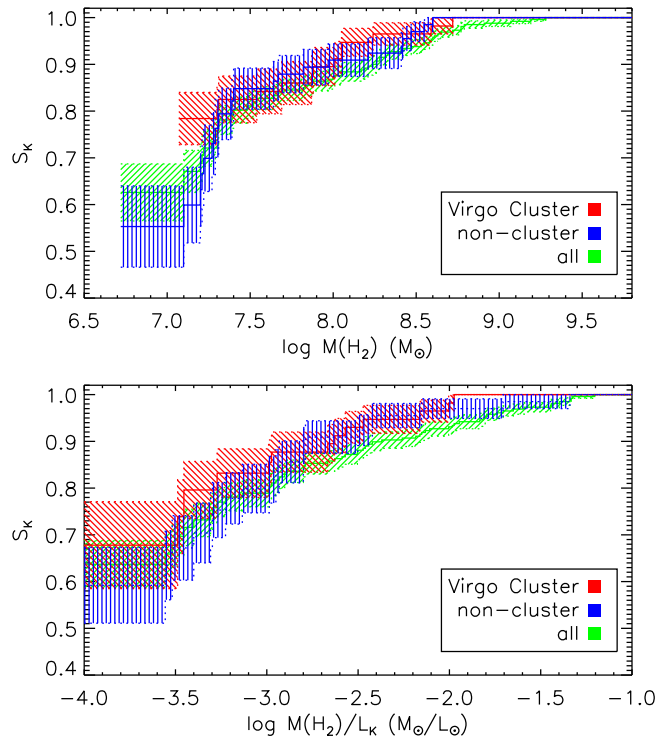}
\caption{The Kaplan-Meier estimators for the cumulative \mhtoo\ and \mhtoolk\ distribution
functions $(S_K)$ for \atlas\ Virgo Cluster members, galaxies outside of the
Virgo Cluster (but at distances $D \leq 24$~Mpc from Earth), and for all \atlas\
members.  The
shaded regions indicate $1\sigma$ uncertainties on the estimated distribution
functions.}
\label{km1}
\end{center}
\end{figure}

The preceding evidence shows that the CO detection rate is nearly the same for
early-type galaxies in the Virgo Cluster and outside the cluster.  
Clearly the galaxies in the cluster are not entirely devoid of molecular gas.  However, 
it is still possible that they may have suffered some modest stripping, which would
show up as systematically smaller \htoo\ masses or \mhtoolk\ ratios for cluster
members than for non-members.  In order to test this hypothesis,
Figure \ref{km1} presents the Kaplan-Meier estimators for the cumulative
\htoo\ and \mhtoolk\ distribution functions for galaxies inside the Virgo Cluster,
galaxies outside the cluster but closer than 24 Mpc, and the entire \atlas\
sample.  In these analyses the null hypothesis is that the cluster and non-cluster
galaxies have the same underlying \mhtoo\ and \mhtoolk\ distributions, and 
a test statistic as large as the observed one would
be expected 25\% to 60\% of the time for \mhtoo\ and 46\% to 57\% of the time for
\mhtoolk.  These probabilities are not so small that the null hypothesis can
be confidently excluded.
In other words, the available data do not
show a striking or reliable difference between the \mhtoo\ and \mhtoolk\
distribution functions for cluster and non-cluster galaxies.

While there may be a slight decrease in the CO detection rate in the Virgo Cluster
compared to field early-type galaxies within 24 Mpc, there is as yet no good evidence that
the molecular gas contents are systematically smaller inside the cluster.
This result stands in opposition to the situation for atomic gas;  HI detection
rates for early-type galaxies in the Virgo Cluster are acutely low compared to
detection rates in the field \citep{alfalfa1,alfalfa2,oo10}.  Additional discussion
of galaxy evolution in the cluster can be found in Section~\ref{sec:discussion}.

\begin{figure*} % velocities in virgo cluster
\begin{center}
\includegraphics[width=16cm,trim=2cm 0.8cm 1cm 1cm]{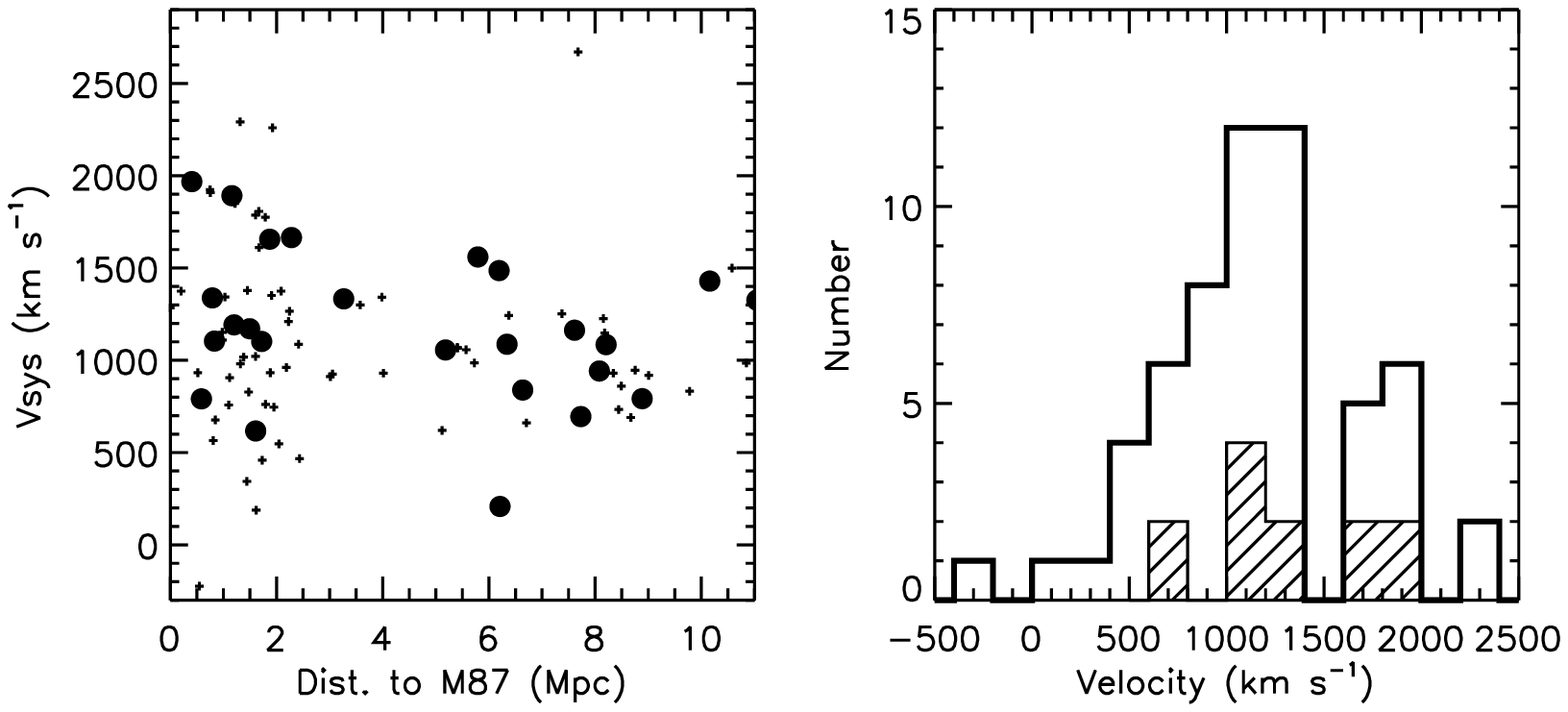}
\caption{Systemic velocities of CO-detected and non-detected galaxies near the
Virgo Cluster.  Symbols in the right-hand plot are as in Figure~\ref{fig:correlations}.
The right panel shows systemic velocities of all \atlas\ Virgo cluster members in the
open histogram (heavy line) and systemic velocities of the CO detections in
the hashed histogram.  Unlike the previous histograms, in this case the vertical
axes for detections and non-detections are not scaled relative to each
other.}\label{fig:virgostats2}
\end{center}
\end{figure*}

Figure \ref{fig:virgostats2} shows the systemic velocities of CO detections
and non-detections as a function of projected distance to M87, as well as the velocity
distributions of \atlas\ Virgo Cluster members and CO detections.  
The Virgo Cluster members have a centrally peaked distribution with a
velocity dispersion of 525 \kms, in good agreement with the value of 573 \kms\
measured for elliptical and lenticular galaxies by \citet{BTS}.  The 12 Virgo
Cluster CO detections have a dispersion of 414 \kms\ and are also peaked
about the cluster systemic velocity.  This behavior is in marked contrast to the 
broad, flat velocity distribution of the spirals in the cluster, which are
characterized by a dispersion of 888 \kms\ \citep{BTS}. 
A more complete characterization of the spirals is provided by \citet{RWK99},
who find that the disturbed spirals are approaching a
relaxed velocity distribution whereas the regular, undisturbed spirals have a 
broad, flat, non-virialized velocity distribution.

\begin{figure}  % LOSVD of Virgo
\begin{center}
\includegraphics[width=8cm,trim=1.5cm 1cm 1cm 0.5cm]{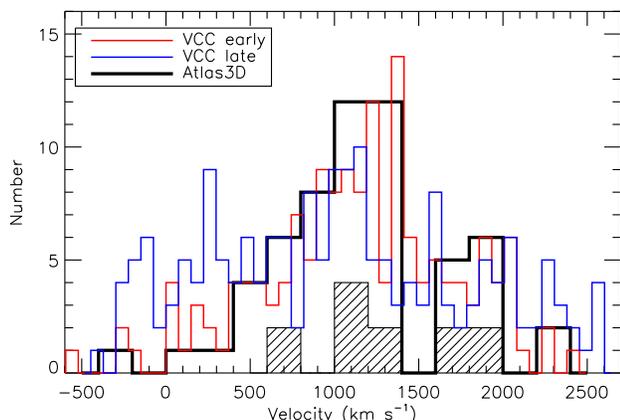} 
\caption{Velocity distributions for \atlas\ Virgo Cluster members, VCC early-type
galaxies, and VCC late-type galaxies.
The \atlas\ CO detections are indicated in the hashed histogram.  \atlas\ galaxies
are binned to 200 \kms; VCC galaxies have smaller bins whose sizes are adjusted
so that their histograms occupy the same area as that of the \atlas\ Virgo Cluster
members.}
\label{VirgoLOSVD}
\end{center}
\end{figure}

The dynamical status of the CO-detected Virgo Cluster early-type galaxies can be
further assessed by
comparing their velocity distribution to those of the other early-type galaxies
and the late-type galaxies in the cluster.  A comparison sample of late-type
galaxies in the Virgo Cluster is provided by the Sa-Sm, Im, BCD, Sp, and dS
galaxies in the Virgo Cluster Catalog \citep[VCC;][]{BST}.  For this purpose we
select from the VCC only the galaxies identified as cluster members and having measured
systemic velocities.   The corresponding early-types are the
VCC E, S0, dE, and dS0.  Velocity distributions for these samples are shown in
Figure~\ref{VirgoLOSVD}.  The results of KS tests on the 
velocity distributions of these various samples are shown in Table~\ref{tab:virgoks}.
Notably, the velocity distribution of the \atlas\ Virgo Cluster CO detections is
consistent with those of the \atlas\ Virgo Cluster CO non-detections and the VCC
early-types.  However, our \atlas\ cluster CO detections are {\it not} consistent with
the VCC late-type galaxies; that null hypothesis can be rejected at the 95\%
confidence level. 
Thus, while the number of CO detections among the \atlas\ Virgo
Cluster galaxies is still relatively small, the indications are that the dynamical
properties of the CO-rich early-type galaxies
are more consistent with the virialized early-type galaxies than with the spirals
and irregulars.  It thus appears unlikely that our CO detections are in new or
recently added cluster members.

%% table 5: Virgo velocity KS tests
%
\begin{table}
\caption{Kolmogorov--Smirnov tests \label{tab:virgoks}}
\begin{tabular}{ l c c}
\hline
Sample     &     A3D Virgo Cluster     &   A3D Virgo Cluster     \\
           &        (all)          &     (CO detections)       \\
\hline
VCC late-types &     0.021          &      0.052    \\
VCC early-types &    0.863          &      0.262    \\
A3D CO nondet. &     0.999        &      0.228    \\
\hline
\end{tabular}
\textit{Notes---} Matrix of probability values for the KS test statistic on the systemic
velocity distributions of \atlas\ galaxies and members of the Virgo Cluster Catalog.
\end{table}
 
\begin{figure} % xrays
\begin{center}
\includegraphics[width=8cm,trim=0.5cm 0.5cm 0.5cm 0.5cm]{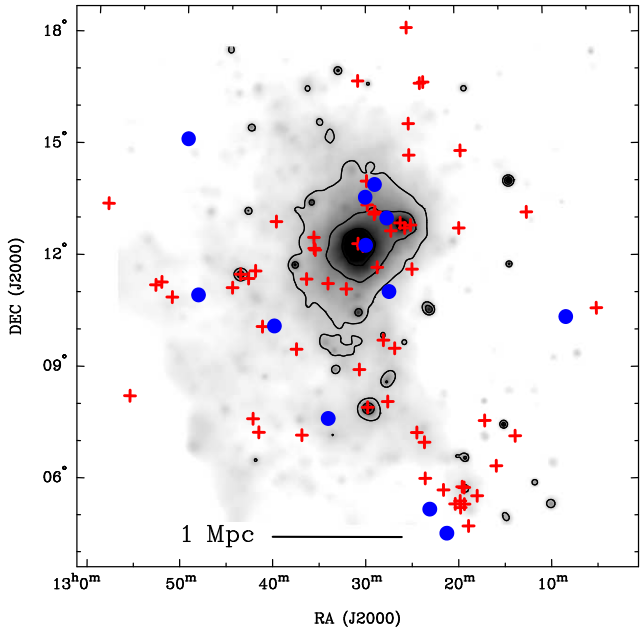}
\caption{CO detections and non-detections towards the centre of the Virgo Cluster.  The greyscale and contours
are the
ROSAT X-ray image of \citet{bohringer}.  The red crosses are the \atlas\ non-detections and
blue filled circles are the CO detections.  The CO detection nearest the X-ray peak is
NGC\,4476, whose regular and apparently undisturbed CO disc was mapped by
\citet{y02}.  
}
\label{fig:xray}
\end{center}
\end{figure}

Figure \ref{fig:xray} shows the spatial distribution of \atlas\ targets and CO
detections in the Virgo cluster, compared to the location of the hot gas
\citep{bohringer}.
Several of the CO detections are seen in projection in the central regions
occupied by hot gas.
Notable among these are NGC\,4429, NGC\,4476, NGC\,4435, NGC\,4477, and NGC\,4459,
all of which are secure CO detections and are within 
2\degr\ (0.6 Mpc) of M87.  NGC\,4435
and NGC\,4477 are members of the Markarian chain to the northwest of M87.
NGC\,4435, NGC\,4476, and NGC\,4459 have distance measurements from 
surface brightness fluctuations \citep{mei07}, and these place them within a true
distance of 1.3 Mpc from M87.
Thus, they must be physically located within the
hot intracluster medium.  In combination with the centrally peaked velocity 
distributions, these data suggest that the CO detections in the 
Virgo Cluster are indeed a dynamically relaxed population that does not avoid the
center of the cluster.  This behaviour is markedly different to that of atomic
gas in the spirals of the Virgo Cluster, which show strong HI deficiencies 
near the center of the Cluster \citep{cayatte90,chung09}.

\subsection{Groups and low density regions}

Figure \ref{fig:virgostats} shows possible evidence for an enhanced CO
detection rate at the outskirts of the Virgo Cluster, $\approx$ 15--20\degr\
(5--8 Mpc) from M87.  This result is driven by the galaxies near NGC\,4278 (north of M87), the
Leo~II or NGC\,3607 group (northwest of M87), and galaxies near NGC\,4684
(south of M87).  
For the galaxies 5 to 8 Mpc from M87 the CO detection
rate is 8/17 = 0.47 \error\ 0.12, not quite $2\sigma$ above the global detection rate.

Paper~VII also presents measurements of a local density $\Sigma_3$ for the \atlas\
members; this estimate gives the projected density of galaxies within a cylinder of
height 600 \kms\ and radius such that 3 neighbors are inside.  As discussed in
Paper~VII, $\Sigma_3$ is large both for members of the Virgo Cluster and for galaxies
in small, tight groups.  A KS test on the $\Sigma_3$ distributions of CO detections
and nondetections (at distances less than 24 Mpc) gives a probability of 7\% that the
two distributions could have been drawn from the same parent -- not a
compelling difference, as we found for the cluster vs. non-cluster comparisons in
the previous section.  
Future studies investigating the effect of
the group environment on CO content or pre-processing of molecular gas in galaxies
infalling into the cluster will require larger samples.

A strong trend in CO detection rate with local density can only be found when
considering the lowest density environments and lowest mass galaxies in this sample.  
The CO detection rate rises to 0.55 \error\ 0.09 for galaxies of $\sigma_e < 100$
\kms\ and density $\log (\rho_{10}/$Mpc$^{-3}) < -1.7$, or 0.41 \error 0.11 for
galaxies of dynamical mass $ < 2\times 10^{10}$ \solmass\ in the same
density regime.  However, as Figure \ref{fig:virgostats} suggests, the lowest
density environments primarily occur at large distances in our sample and they drop
out of the subsamples limited to 24 Mpc.  In addition, the trend with local density is
not significant for galaxies of $\sigma_e > 100$ \kms\ or mass $ > 2\times 10^{10}$
\solmass. Thus, this increase in the CO detection rate of low mass, relatively
isolated early-type galaxies did not manifest itself in the previous discussions of
environmental effects.

The half-dozen most CO-rich galaxies of the sample do tend to be in relatively low density
environments, which is undoubtedly an important clue to their formation histories.
This CO-rich subsample includes NGC\,1266, NGC\,2764, NGC\,1222, NGC\,3665, UGC\,09519, and
NGC\,6014.  
These galaxies are all more distant than 20 Mpc, in regions where the local
luminosity density is a factor of 10 to 200 lower than that in the Virgo cluster,
and the distances to their
nearest $M_K < -21.5$ neighbors are $\ge$ 1 Mpc.
If the accretion of cold gas is most efficient in these kinds of poor environments, 
such accretion might provide a natural explanation for the most CO-rich \atlas\
galaxies.  Signatures of the accretion could be sought in the CO kinematics (through
interferometric imaging) or in the gas-phase metallicity.

\subsection{Young stellar populations}

As one might expect, the molecular gas content of early-type galaxies is tightly
correlated with dust and young stellar populations, which are indicated for the
\atlas\ sample in Paper~II.  
The CO detection rate is 37/48 (0.77 \error\ 0.06)
in galaxies with dust discs, dusty filaments, or blue regions, but only 17/208 (0.08 \error\ 0.02) in
galaxies with none of those features.

\begin{figure} % hbeta
\begin{center}
\includegraphics[width=8cm,trim=0.8cm 0.5cm 0.5cm 0.5cm]{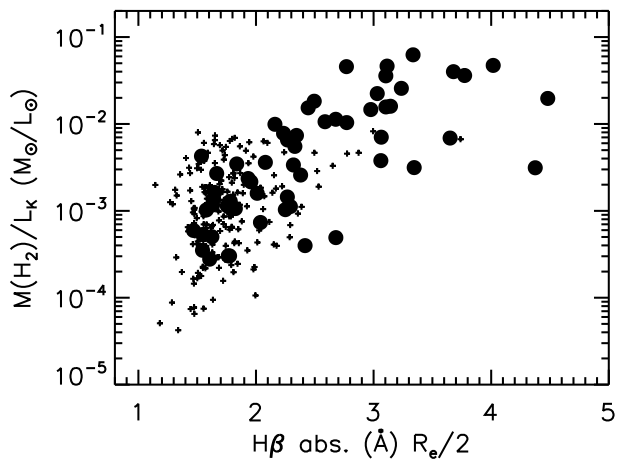}
\caption{Molecular mass and H$\beta$ absorption line strength (McDermid et al.\ in
prep).  Symbols are as in
Figure~\ref{fig:correlations}.   
}
\label{fig:hbeta}
\end{center}
\end{figure}

Figure \ref{fig:hbeta} shows that the galaxies that are rich in molecular gas
also have evidence of young stars, as traced by the H$\beta$
absorption line strength index in the classic Lick/IDS system from
our SAURON spectroscopy (McDermid et al.\ in prep).   Here the H$\beta$ absorption line strength
is measured in an aperture of radius $R_e/2,$ which has a median value of
9.3\arcsec\ for the sample so is similar to the spatial region in which we have
searched for CO J=1-0 emission.  However, the trend is not qualitatively different
when apertures of radius $R_e$ or $R_e/8$ are used.  Naturally, as H$\beta$ is an equivalent
width measure, both it and \mhtoolk\ are roughly normalized to the total stellar
luminosity.

There is a clear trend for galaxies with strong H$\beta$ absorption to exhibit
higher \mhtoolk\ ratios, indicating that higher gas fractions are found
in galaxies with larger proportions of young stars.  Intuitively, the trend
makes sense, since the \htoo\ reservoir is thought to fuel any ongoing
star-formation.  However, the relation is not especially tight, with more than an
order of magnitude variation in \mhtoolk\ at a given line strength.  This spread
probably reflects
the differences between these two tracers. The mass of \htoo\ should trace
the {\it instantaneous} star-formation activity, with some caveats noted below; 
H$\beta$, in combination with metal indices, traces a weighted average age which is
strongly biased towards recent star formation activity.
Differences in recent star formation histories (especially bursty or
merger-driven activity) could produce some of the scatter.
Some scatter is also probably due to the fact that H$\beta$ is degenerate
for a broad range of possible star formation histories and does not exclusively
trace recent ($< 1$ Gyr) events.

Another cause of scatter in the H$\beta$-\htoo\ trend
could be differing star formation
efficiencies, driven by several mechanisms.  For
example, the shape and depth of the potential can affect the stability of the cold gas
disc embedded within it \citep{martig}.  It may also be that the
molecular gas reservoir is prevented from forming stars in some galaxies due
to some form of heating, e.g.\ from AGN activity or hot phases of stellar
evolution.  \citet{crocker-all} explore the latter in some detail, proposing
connections to the optical emission from the ionized gas.  We postpone such
analysis using the full \atlas\ data complement for future papers in this
series.

There are a number of CO upper limits found in galaxies with strong H$\beta$ 
absorption ($> 2.3$ \AA).  At face value, these facts could indicate the
presence of a significant young stellar population without any associated
molecular gas, perhaps immediately following exhaustion of the gas after a
significant star-formation event.  However, those \htoo\ upper limits
are quite high because the galaxies are distant. 
These non-detections may be comparable in their \htoo\ properties to the
detected galaxies, simply falling just below our sensitivity level.
At the other extreme, while there are many objects with intermediate values
of H$\beta$ that are both CO-detected and undetected, there are no CO detections in
objects with H$\beta \leq 1.4$ \AA.  The galaxies showing no evidence for
young stellar populations 
also show no evidence of molecular gas within our detection thresholds.

\section{DISCUSSION}\label{sec:discussion}

\subsection{Rejuvenation}

\citet{thomas10} have used SDSS spectroscopy to
measure luminosity-weighted simple stellar population (SSP)
ages for a sample of low redshift, morphologically selected early-type galaxies.
They interpret a population of younger galaxies as examples of 
a rejuvenation phenomenon driven by recent ($< 2$ Gyr) star formation.  They also
argue that this rejuvenation phenomenon is mass- and density-dependent, being more
common (or perhaps stronger?) in low mass galaxies in low density environments.

This picture is at least qualitatively consistent with what we find for the
molecular gas content of early-type galaxies, with a few important caveats.  For
example, the
results presented above suggest that the environmental density dependence of the 
\htoo\ content is extremely
subtle except at the lowest densities.  We do
detect molecular gas in galaxies covering the full range of densities and dynamical
masses probed in our sample.  Thus, if the rejuvenation episodes are indeed
connected to the present-day \htoo\ content, we should expect that they are not
exclusively limited to low mass galaxies in low density environments.  A model which
fits all of this data
should also allow for significant rejuvenation rates in high mass galaxies and in
high density environments.

\subsection{Molecular masses of early-type galaxies}

The cumulative \htoo\ and \mhtoolk\ distribution functions in Figures~\ref{km2}
and \ref{km1} remind us that the majority of the galaxies studied here are not
detected in CO emission, so that we are really only studying the gas-rich
ends of the distribution functions.  

Given that the galaxies studied here are early-type galaxies, the molecular masses
detected are surprisingly high in some cases.  Three of the sample galaxies have molecular
masses greater than $10^{9}$ \solmass\ (assuming their true CO-\htoo\ conversions are
not too different from that used here).   These galaxies have comparable
amounts of molecular gas to the Milky Way and several times more than 
in M31 \citep{dame93}, yet they are clearly not spirals.  The highest molecular
masses found in this sample are also comparable to that of the merger remnant
NGC\,7252 \citep{dupraz90} and of some lenticular polar ring galaxies
\citep{galletta97}.
Nine of the sample have molecular masses greater than $5\times 10^8$
\solmass, which is comparable to the molecular mass of the radio galaxy Fornax~A
\citep[NGC\,1316;][]{horellou01}.  
The mass of molecular gas in Cen~A (NGC\,5128) is also a few $10^8$
\solmass\ \citep{morganti2010}, 
at an assumed distance of 4.2 Mpc
\citep{tonrySBF}.  These two galaxies meet all of the \atlas\ selection criteria except for
the Declination limit, so they could have been in the sample if they were elsewhere on
the sky.  The comparison to Cen~A is especially notable because, in spite of its most
flamboyant optical appearance, Cen~A would not have been the most CO-rich galaxy in the
sample.  At \mhtoolk\ $ \approx 4\times 10^{-3}$, it would also have been solidly in
the middle of our detected \mhtoolk\ values ($3 \times 10^{-4}$ to $6.3 \times
10^{-2}$ \solmass/\solum).  Evidently the molecular gas content of a galaxy does
not serve as a reliable predictor of its optical morphology, nor does the
optical morphology reliably predict the molecular content.

The molecular gas masses detected here are, however, not quite in the same league as that
of Perseus~A (NGC\,1275), which has some $10^{10}$ \solmass\ of cold \htoo\
\citep{salome06}.  NGC\,1275 is the
central galaxy of the Perseus Cluster, and \citet{lim08} have recently shown that 
its molecular gas is distributed in filaments which are free-falling towards the
center of the potential.  The behaviour is kinematically consistent with the 
molecular gas having cooled and condensed out of the galaxy's X-ray halo.
But this behaviour is probably not taking place in the bulk of the CO-detected \atlas\
galaxies, since the preponderance of double-peaked CO profiles suggests relaxed discs
in regular rotation.  If the molecular gas in the \atlas\ galaxies did originate
through condensation, it must have happened at least a
crossing time ago, so that the gas has settled.

Both the CO detection rate and the \htoo\ mass distribution function 
are independent of the stellar luminosity, for early-type galaxies.
When considering all types of galaxies, of course, there is a very strong
dependence of galaxy properties on the stellar mass; galaxies with stellar masses
above $3\times 10^{10}$ \solmass\ tend to be old spheroids, hence poor in cold gas, 
and less massive
galaxies tend to be HI- and CO-rich young discs \citep[e.g.][]{kauffmann03}.  Among
the red sequence galaxies, \citet{vdwel09} argue that a stellar mass of $10^{11}$
\solmass\ is a watershed above which there are very few flattened, disc-like systems.
However, {\it among the red sequence galaxies}, we find no evidence that a 
stellar mass of $3\times 10^{10}$ \solmass\ or $10^{11}$ \solmass\
has any particular significance for a galaxy's molecular content.

\subsection{On the origin of the molecular gas in early-type galaxies}

Our results show that {\it at least} 22\% of all early-type galaxies contain molecular
gas, and Figure~\ref{km1} presents the inferred \mhtoolk\ distribution for
\mhtoolk~$ > 10^{-4}$ \solmass/\solum.  It is worthwhile to recall that this volume-limited
survey, complete down to $M_K = -21.5$, has no selection on either FIR
luminosity or $B$ magnitude, so that it finally eliminates selection
bias towards the galaxies which are more likely to have star formation
activity.  

Using interferometric HI observations, \citet{oo10} have 
recently argued that ongoing accretion of cold atomic gas is common for field
early-type galaxies.  In some cases the atomic gas retains a large angular momentum
and forms structures which are tens of kpc in extent, but in other cases the atomic
gas is found in the central few kpc of the galaxy, coincident with a molecular disc.
\citet{oo10} thus propose that at least some of the molecular gas in early-type
galaxies may have been accreted in this manner, possibly in the form of atomic gas,
settling into a regular disc and converting into molecular gas as its density
increases.  Future comparisons
of molecular, atomic, ionized gas and stellar kinematics in the \atlas\ galaxies
should help
quantify the role of cold accretion as an origin for the molecular gas.

There is a notable decrease in the CO detection rate of the slow
or ``non-regular'' rotators.  Since this effect 
remains even after controlling for the mass distributions of slow and fast
rotators, it must be driven by $\lr$ rather than by mass.   
In other words, the molecular content is
more strongly correlated with the types of orbits the stars occupy
than with the number of stars.  

The rate of internal stellar mass loss in a galaxy should naturally be independent
of the angular momentum of the stars.  However, the mass loss material should
retain some memory of the specific angular momentum of its progenitor stars.
If the material suffers shock heating, cooling, and condensation to form molecular
gas, it may lose some specific angular momentum in the process, but we would
generally expect the mass loss material in a slowly rotating galaxy to have
smaller specific angular momentum than that in a fast rotator.  If this scenario
accounts for the bulk of the molecular gas in early-type galaxies, then, we might
expect the molecular gas discs in the galaxies of small (intrinsic) $\lr$ to be
more compact than in the large-$\lr$ galaxies.  Indeed, the molecular gas in the
galaxies of small $\lr$ might have dropped to the nucleus and been consumed,
instead of forming a kpc-scale disc.
From a different perspective, if the formation processes of slow rotators are
different from those of fast rotators, it might also be the case that the
formation of a slow rotator inevitably destroys any cold gas disc.  In short,
there are at least two scenarios which can explain a lack of cold gas in the slow
rotators.

The detection of CO in Virgo Cluster early-type galaxies provides a useful
perspective on the issue of the origin of the molecular gas.  We have argued
above that the CO detection rate among \atlas\ Virgo Cluster members may be
modestly lower than the detection rate outside the cluster, but the statistical
significance of the result is not great.  Furthermore, the \mhtoo\ and
\mhtoolk\ distributions inside and outside the cluster are consistent.  We have
also argued that the CO-detected galaxies are virialized in the cluster
potential and deeply embedded within the hot intracluster gas.  Despite a long
residence in the cluster, their molecular content is not much lower than in
early-type galaxies outside the cluster.

Both the internal and external scenarios for the origin of molecular gas seem
significantly more problematic when galaxies are already deep in the cluster
potential.  Cold mode accretion of gas through filaments may be viable for
relatively isolated galaxies, but gas falling into the cluster should more
likely end up as hot gas in the general cluster potential than as cold gas in a specific galaxy.  
Acquisition of molecular gas through major or minor mergers also should be rare,
both because the cluster's high velocity dispersion decreases the merger rate and 
because mergers tend to fling gas to large radii \citep{barnes02} where it would be
highly
vulnerable to stripping.  These general expectations are supported by the
N-body plus hydrodynamical simulations of \citet{Tonnesen07}, which find that
galaxies commonly accrete gas when they are far in the outskirts of a cluster
but rarely accrete gas after they pass through the cluster's virial radius.
In addition, the internal stellar mass loss scenario also seems difficult
because that material is expected to shock heat to X-ray temperatures
\citep[e.g.][]{MB03} and the concomitant low densities will again make the
material vulnerable to stripping.  More detailed simulations of the evolution of
hot gas in cluster members (not cluster-dominant or cluster-centered galaxies)
would help to quantify this impression.  In short, these considerations suggest
that the CO-detected Virgo Cluster early-type galaxies probably did not acquire
their cold gas after entering the cluster.   This in turn suggests that
they have retained their molecular gas since they fell into the cluster, and that
the gas has not cycled through a low density phase in that time.

\subsection{On the removal of \htoo}

If the CO-rich Virgo Cluster early-type galaxies are indeed virialized, so that
they have been in the cluster for at least a relaxation time yet have still
retained their molecular gas, we conclude that it must be difficult to remove 
the molecular gas entirely.    For a cluster such as Virgo, with several hundreds
to thousands of galaxies, the relaxation time is at least 10 crossing times
\citep{BT} and the crossing times are on the order of a Gyr.  Thus we
infer that the virialized, CO-rich early-type galaxies have probably retained
their gas over at least several Gyr in the cluster.

It has been suspected for many years that it would be more difficult to remove
the molecular gas from cluster galaxies than to remove the atomic gas.
The reasoning is straightforward: the molecular gas has much higher volume and
surface densities than atomic gas, and in addition it tends to be deeper in
the galaxy's potential well, both of which make it more tightly bound.  
Individual galaxies do sometimes show
evidence that their molecular hydrogen is being stripped \citep{Siva10,vollmer08},
but in general the observations of molecular gas in Virgo Cluster spirals have
suggested that \htoo\ deficiencies are subtle, even when HI deficiencies are
strong \citep{ky89}.  The same now appears to be true for Virgo Cluster early-type
galaxies, where  \htoo\ deficiencies are subtle to nonexistent
whereas HI deficiencies are strong \citep{alfalfa1,alfalfa2}.
In this context, the Virgo Cluster early-type galaxies provide a useful 
perspective because (as we argued above) they have
been residing in the cluster longer than the spirals.  They therefore provide longer
time-baseline views into the gas removal processes than spirals do.  

Recent adaptive-mesh hydrodynamical
simulations of \citet{tonnesen09} suggest that when a galaxy first falls
into the intracluster medium, it suffers a brief but intense period of stripping
when most of the loosely-bound material is removed.  In the simulations, 
dense molecular gas can
be retained through this period, in agreement with the implications of our work.
Quantitatively, the standard \citet{gunngott} stripping criterion
requires a ram pressure $\sim V_c^2 \Sigma_{gas}/R.$  
For typical values we take a circular velocity $V_c$ of 200 \kms, 
molecular gas surface density $\Sigma_{gas}$ at 100 \msunsqpc, 
and a radius $R$ of 1 kpc, based on the interferometric CO
observations of \citet{crocker-all}.  The ram pressure needed is then $\sim 3\times 10^{-9}$ 
dyne~cm$^{-2}$.  \citet{vollmer09} finds that these kinds of pressures require a galaxy
to approach within $\sim 50$ kpc of the Virgo Cluster center even if the velocity at
closest approach is 2000 \kms.  Thus the molecular gas should be retained except on
very high eccentricity orbits within the cluster.
\citet{tonnesen09} also mention that angular
momentum transfer to the non-rotating cluster gas can gradually drive the
remaining molecular 
gas towards the center of its host galaxy.  For all of these reasons, the CO-rich Virgo Cluster 
early-type galaxies could motivate longer simulation runs in an attempt to test
whether these processes can transform an exponential gas disc (typical of
spirals) into a rather compact, sharp-edged molecular disc such as those observed
in NGC\,4459, NGC\,4526, and NGC\,4477 \citep{YBC, crocker-all}.

\subsection{On the formation of fast rotators}

While our early-type galaxies are today rather red in colour, they must have been
blue during the epoch when their stars were forming.  The mass distributions and
morphologies of their progenitors are, however, the subjects of much debate.
One major question is whether or not the fast rotator galaxies 
passed through a stage as spiral galaxies.  Some authors suggest not.  For
example, \citet{dekel09} propose that if the matter accreting through cold streams
into a protogalactic
halo is highly clumped, the dynamical interactions between the
clumps will counteract the tendency of cold gas to settle into a disc.  In this
manner a spheroid-dominated galaxy could form without ever passing through a
disc-dominated stage, even if the bulk of the accreted baryons are in the form of
cold gas.

Other popular scenarios propose to form fast rotator early-type galaxies 
by converting spirals.
One aspect of the conversion must involve quenching the
star formation activity (so as to redden and age the stellar population);
another aspect is to effect the morphological change 
(dispersal of the spiral structure, disc thickening and/or bulge growth).  
It is often assumed that the removal of the
cold gas will accomplish these objectives.  
Turning off new star formation would remove the continuous input of a dynamically cold
stellar population, so that dynamical heating and thickening would naturally
erase the non-axisymmetric spiral structures.  
Removal of the cold gas may not be necessary for this conversion
process, though.
\citet{bournaud07} have suggested that multiple minor
mergers (with mass ratios of 10:1 and greater) will gradually thicken a spiral's stellar
disc and increase its central concentration (Sersic index).
\citet{martig} also remind us that if the gravitational potential can be
sufficiently deepened, a cold gas disc could be simply rendered incapable of star
formation activity.
Our current work suggests that, indeed, 
some cluster galaxies can retain significant amounts of molecular gas even after
being resident in the cluster for several Gyr.

We cannot, at present, confirm or refute the speculation that fast rotator
early-type galaxies in the Virgo Cluster are the stripped,
heated, and/or quenched remnants of spirals.
We simply comment that {\it if this is the case,} the paradigm should not
necessarily require all of the molecular gas to be removed in the conversion
process. 
In a purely hydrodynamic stripping scenario, it would be interesting to test
whether the star formation
quenching and morphological change can be accomplished quickly enough by removing
only {\it some} of the
molecular gas and simply rearranging the rest.  In the merger, morphological
quenching, and clumpy accretion
scenarios, it would
be valuable to have statistical predictions for the properties (masses and spatial
extents) of any cold gas discs which might remain.

\section{SUMMARY}
\label{sec:summary}

We present a $^{12}$CO J=1-0 and J=2-1 search in the galaxies of the \atlas\ sample,
a complete volume-limited sample of early-type galaxies with $-21.5 \geq M_K \geq
-26$ and with morphologies 
verified by inspection of optical (mostly SDSS) images.  CO data from the IRAM 30m
telescope are collected for 259 of the 260 members, including 204 new observations
and the remainder collected from the literature.
The $3\sigma$ upper limit for a sum over a 300 \kms\ linewidth corresponds to an
\htoo\
mass  $\sim 1\times 10^7$~\solmass\ for the nearby sample members (11 Mpc) and a mass
$\sim 1\times 10^8$~\solmass\ for the most distant members (40 Mpc).  A few targets
are observed significantly more deeply than that.  

The detected CO line intensities correspond to \htoo\ masses of $\log
\rm M(H_2)/M_\odot$ = 7.10 to 9.29 and \mhtoolk = $2.8\times 10^{-4}$ to 0.063
\solmass/\solum.
The CO detection rate is 56/259 = 0.22\error 0.03, a rate
known to be a lower limit since (1) molecular gas outside the
central 22\asec\ has not been observed, (2) we expect to be incomplete for linewidths
greater than 300 \kms, and (3) compact
distributions suffer strong beam dilution.  Many of the detections are brighter in
CO(2-1) than in CO(1-0).  The line profiles are often double-horned and/or
asymmetric.  These observations collectively suggest that the molecular gas is often found in 
regular rotating discs.   It is sometimes in compact structures
(smaller than the 12\asec\ CO(2-1) beam, which is 1.1 kpc at 20 Mpc) and sometimes in
structures comparable to or larger than the CO(1-0) beam (2.2 kpc at 20 Mpc).
The linewidth distribution peaks at smaller widths than expected, suggesting that
we have some cases in which the molecular gas reaches the galaxy's asymptotic
circular velocity and some in which it doesn't.
Molecular surface densities, averaged over the inner 2 to 4 kpc of the
galaxies, are comparable to those found in spirals.
Interferometric
observations will be necessary in order to quantify those general impressions
about the typical gas distributions.

There is a strong correlation between the presence of molecular gas and the
presence of dust, blue features, and young stellar ages seen in H$\beta$
absorption.  Thus, the molecular gas that we detect is often engaged in star
formation.

The CO detection rate is not a function of luminosity over the range $-21.5 \geq
M_K \geq -26$, and the \htoo\ mass distribution function is also independent of
stellar luminosity.  The CO detection rate is a strong function of the global
velocity width \sigmae, with galaxies of $\sigma_e < 100$ \kms\ more than twice as
likely to be detected.  However, the dependence on the dynamical mass inferred from
modelling the stellar kinematics is considerably more subtle.  These considerations
suggest that the \sigmae\ effect is probably a combination of a downsizing trend
and an observational bias in favor of detecting CO in face-on galaxies.  

The CO detection rate is strongly correlated with the stellar specific angular
momentum and internal kinematic substructure, such that slow rotators and
non-regular rotators have much less molecular gas than their fast and regular
rotating counterparts.  This result remains true after controlling for the \sigmae\
and dynamical mass dependences noted above.  We suggest that either the assembly
processes which built the slow rotators also destroyed or removed their molecular
gas, or they are less likely to re-acquire molecular gas after assembly.

Approximately half of the fast rotators at distances less than 24 Mpc are members
of the Virgo Cluster and half are not.  The CO detection rate is modestly lower in
cluster members than in non-members; however, it is important to note that
the drop in detection rate is only a $\sim 1\sigma$ effect.
There is no measurable difference in the \htoo\ or \mhtoolk\ distribution functions
of cluster and non-cluster galaxies within 24 Mpc.  CO-rich Virgo Cluster
early-type galaxies are virialized in the cluster potential and they do not avoid
the center of the cluster (unlike cluster spirals).  Apparently they have retained their molecular gas
through extended residences in the cluster.

The rarest, most gas-rich galaxies (\mhtoolk $> 0.02$ \solmass/\solum) are found 
in relatively low-density environments at distances beyond 24 Mpc.
These results suggest that the CO-detected Virgo Cluster galaxies have CO-detected
analogs of the same \mhtoolk\ in the field, but there is in addition a 
small population
of CO-rich objects in the field that is not found in the Virgo Cluster.

\section*{Acknowledgments}
Thanks to Aeree Chung for providing the ROSAT Virgo cluster image in FITS format.
The research was partially supported by grant NSF-0507432 to LMY, who also
gives hearty thanks to the University of Oxford sub-department of
astrophysics for hospitality under the STFC visitor grant.
MC acknowledges support from a STFC Advanced Fellowship (PP/D005574/1) and a
Royal Society University Research Fellowship.
This work was supported by the rolling grants `Astrophysics at Oxford' PP/E001114/1
and ST/H002456/1 and visitors grants PPA/V/S/2002/00553, PP/E001564/1 and ST/H504862/1
from the UK Research Councils. RLD acknowledges travel and computer grants from Christ
Church, Oxford and support from the Royal Society in the form of a Wolfson Merit Award
502011.K502/jd. RLD also acknowledges the support of the ESO Visitor Programme which
funded a 3 month stay in 2010.
SK acknowledges support from the the Royal Society Joint Projects Grant JP0869822.
RMcD is supported by the Gemini Observatory, which is operated by the Association of
Universities for Research in Astronomy, Inc., on behalf of the international Gemini
partnership of Argentina, Australia, Brazil, Canada, Chile, the United Kingdom, and
the United States of America.
TN and MBois acknowledge support from the DFG Cluster of Excellence `Origin and
Structure of the Universe'.
MS acknowledges support from a STFC Advanced Fellowship ST/F009186/1.
NS and TD acknowledge support from an STFC studentship.
The authors acknowledge financial support from ESO.
LMY was assisted by statistician Marge Inovera, courtesy of T.\ and R.\ Magliozzi.

\clearpage

% Table 1: rms noises and integrated intensities (for all our data)
%
\begin{deluxetable}{lccccccccc}
\tablewidth{0pt}
\tablecaption{CO intensities and \htoo\ masses. \label{tab:intensities}}
\tablehead{
\colhead{Galaxy} & \colhead{rms (1-0)} & \colhead{rms (2-1)} &
\multicolumn{2}{c}{Vel. range} & \colhead{$I_{\rm (1-0)}$}
& \colhead{$I_{\rm (2-1)}$} & \colhead{Ratio} & \colhead{Source} & \colhead{log M(H$_2$)} \\
\colhead{} & \colhead{mK} & \colhead{mK} & \multicolumn{2}{c}{\kms} &
\colhead{\kkms} & \colhead{\kkms} & \colhead{(2-1)/(1-0)} & \colhead{} &
\colhead{\solmass} \\ }
\tablecolumns{10}
\startdata
% Mon Feb  7 13:49:28 2011  Using optical velocities from our new data.
%       Name      rms(1-0)     rms(2-1)       Start       End           Int(1-0)       Int(2-1)       Ratio   Source   M(H2)
%                 mK, Tmb      mK             km/s        km/s          K km/s         K km/s                           Msun
% 
      IC0560 &      3.86 &      9.40 &   1703.0 &   2003.0 &     0.106 ( 0.347) &    -0.830 ( 0.758) & \nodata &     1 &   $<$  7.67 \\
      IC0598 &      4.19 &      5.72 &   2106.0 &   2406.0 &    -0.007 ( 0.457) &    -0.340 ( 0.525) & \nodata &     1 &   $<$  8.02 \\
      IC0676 &      4.84 &      4.92 &   1310.0 &   1516.0 &    11.641 ( 0.429) &    16.654 ( 0.343) &   1.43 (0.06) &     1 &     8.63 (0.02) \\
      IC0719 &      2.80 &      2.74 &   1619.0 &   2022.0 &     3.472 ( 0.361) &     5.798 ( 0.311) &   1.67 (0.20) &     1 &     8.26 (0.04) \\
      IC0782 &      3.05 &      3.86 &   2274.0 &   2574.0 &     0.337 ( 0.343) &     0.072 ( 0.333) & \nodata &     1 &   $<$  7.92 \\
      IC1024 &      4.75 &      6.35 &   1359.0 &   1619.0 &    11.419 ( 0.428) &    15.631 ( 0.401) &   1.37 (0.06) &     1 &     8.61 (0.02) \\
      IC3631 &      3.23 &      3.17 &   2672.0 &   2972.0 &     0.213 ( 0.268) &    -0.006 ( 0.235) & \nodata &     1 &   $<$  7.94 \\
     NGC0448 &      3.72 &      4.57 &   1758.0 &   2058.0 &    -0.399 ( 0.345) &    -0.767 ( 0.395) & \nodata &     1 &   $<$  7.74 \\
\enddata
\tablecomments{Intensities are quoted as main beam brightness temperatures $T_{mb}$.  
The rms noise values are for channels binned to 31 \kms.
The velocity range is either 300 \kms\ (for nondetections) or the actual range used
in the sums (for detections).
All quoted uncertainties are $1\sigma$; \htoo\ mass upper limits are $3\sigma$.  
Where the
line ratio is a limit, it is computed using the $3\sigma$ upper limit on the
undetected line.  
\htoo\ masses assume the distances listed in \citet{paper1}.
For galaxies taken from the literature, we give only the \htoo\ mass corrected to
our assumed distance and other quantities may be found in the original sources.
Data sources: 1 = this paper, 2 = \citet{CYB}, 3 = \citet{WS03}, 4 =
\citet{SWY1}, 5 = \citet{wch95}, 6 = \citet{y02}, 7 = \citet{WSY10}.
The table is available in its entirety from the project website
http://purl.com/atlas3d.
}
\end{deluxetable}

\clearpage

% Table 3:  CO line parameters (detections only)
\begin{deluxetable}{lccccccccl}
\rotate
\tablewidth{0pt}
\tabletypesize{\footnotesize}
\tablecaption{Fitted properties of detected lines \label{tab:detections}}
\tablehead{
\colhead{Galaxy} & \colhead{$I_{\rm (1-0)}$} & \colhead{$I_{\rm (2-1)}$} &
\colhead{$V_{\rm (1-0)}$} & \colhead{$V_{\rm (2-1)}$} & \colhead{$\Delta
V_{\rm (1-0)}$} & \colhead{$\Delta V_{\rm (2-1)}$} & 
\colhead{$T_{\rm mb\,(1-0)}$} & \colhead{$T_{\rm mb\,(2-1)}$} & \colhead{Comments}\\
\colhead{}  & \colhead{K km s$^{-1}$} & \colhead{K km s$^{-1}$} &
\colhead{km s$^{-1}$}   & \colhead{km s$^{-1}$}   & \colhead{km
s$^{-1}$}  & \colhead{km s$^{-1}$}  &  \colhead{mK} & \colhead{mK} & \colhead{}}
\tablecolumns{8}
\startdata
% Mon Dec 29 20:30:54 2008
%    Name           Area10             Area21             Vsys10            Vsys21             FWHM10               FWHM21          Tmb10     Tmb21       Comments
%                   K km/s             K km/s             km/s              km/s               km/s                 km/s            mK        mK 
% 
      IC0676 &   12.11 (  0.40) &   17.03 (  0.33) &    1429.0 ( 1.6) &    1428.0 ( 0.8) &      97.4 ( 3.4) &      87.1 ( 1.7) &   116.8 &    184.3 & \\
      IC0719 &    3.77 (  0.37) &    6.10 (  0.33) &    1859.0 (14.8) &    1836.0 ( 6.8) &     292.6 (29.0) &     267.3 (14.1) &    12.1 &     21.5 & \\
      IC1024 &   12.03 (  0.44) &   16.24 (  0.43) &    1466.0 ( 3.0) &    1492.0 ( 1.9) &     160.9 ( 5.9) &     146.9 ( 3.8) &    70.3 &    103.9 & a; 2h? \\ %st sides; asymm; width + center OK\\
     NGC0509 &    0.49 (  0.15) &    \nodata       &    2288.0 ( 5.9) &     \nodata      &      35.4 (11.6) &       \nodata    &    13.0 &  \nodata &  \\ %OK\\
     NGC0524 &    3.11 (  0.36) &    5.30 (  0.52) &    2393.0 (21.2) &    2412.0 (17.1) &     341.2 (37.0) &     336.6 (34.8) &     8.6 &     14.8 & 2h? \\
     NGC1222 &   18.36 (  0.41) &   29.41 (  0.47) &    2410.0 ( 1.6) &    2399.0 ( 1.0) &     142.4 ( 3.3) &     132.6 ( 2.2) &   130.  &    237.  & a* \\ %quite asymm do peak by hand\\
     NGC1266 &   32.30 (  0.47) &   70.54 (  2.50) &    2161.4 ( 0.9) &    2161.9 ( 0.4) &     129.4 ( 2.3) &     127.2 ( 2.0) &   234.5 &    520.8 & \\ %broad wings in 2-1\\
             &    \nodata       &   36.92 (  2.27) &       \nodata    &    2151.9 ( 4.3) &       \nodata    &     349.9 (18.3) & \nodata &     99.1 & wings \\
\enddata
\tablecomments{
Columns: (1) Source name; 
(2)--(3) line area in CO(1-0) and CO(2-1); 
(4)--(5) central velocity in CO(1-0) and CO(2-1); 
(6)--(7) FWHM in CO(1-0) and CO(2-1); 
(8)--(9) peak brightness temperature in CO(1-0) and
CO(2-1). Formal 1$\sigma$ uncertainties in each quantity are quoted in parentheses. 
Comment `a' in column (10) indicates the line profile is notably
asymmetric, and `a*' indicates the line is sufficiently asymmetric that the
peak brightness temperature (and area, systemic velocity and width, for NGC\,3665 in 2-1) has been measured directly off the profile rather than
with a fit.  Comment `2h' indicates a double-horned
line profile.  
Notes on specific galaxies:
{\bf NGC\,1266} - the 1-0 line is, formally speaking, adequately fit with one Gaussian
component but the 2-1 line has broad, low-level wings which require two Gaussian
components of the same center and different widths.
{\bf NGC\,6014} - the line profile is neither a double horn nor a Gaussian, but the
central dip suggests a ringlike CO distribution.
{\bf NGC\,4281, 5638 and 7454} - the spectrum shows hints of two peaks.
The peaks were fitted independently and both velocities are listed.
Formally the Gaussian fit detects each of these with an integrated area 
greater than 3$\sigma$, but as there is no emission within 300 \kms\ of the
optical velocity it is a nondetection in Table~\ref{tab:intensities} and it is NOT
counted as a detection the rest of the analysis.
{\bf NGC\,4643} is formally detected via the integrated intensity in the central 300
\kms\ of the 2-1 spectrum and its line is also suggestive of two peaks.  
As for NGC\,4281, the two velocities listed are those of the two peaks (rather than the
systemic velocity that one would infer from assuming those are two horns).
}
\end{deluxetable}

%----------  THE END
\end{document}